\documentclass[]{pasj01}
\draft

\begin{document} 
\Received{2017/05/18}
\Accepted{}

\title{Energy Dependent Intensity Variation of the Persistent X-ray Emission of Magnetars Observed with Suzaku}

%
\author{Yujin \textsc{Nakagawa}\altaffilmark{1}, Ken \textsc{Ebisawa}\altaffilmark{2} and Teruaki \textsc{Enoto}\altaffilmark{3}}
\altaffiltext{1}{Center for Earth Information Science and Technology, Japan Agency for Marine-Earth Science and Technology,
3173-25 Showa-machi, Kanazawa-ku, Yokohama, Kanagawa 236-0001, Japan}
\altaffiltext{2}{Institute of Space and Astronautical Science, Japan Aerospace Exploration Agency,
3-1-1 Yoshinodai, Chuo-ku, Sagamihara, Kanagawa 252-5210, Japan}
\altaffiltext{3}{The Hakubi Center, Kyoto University, Yoshida-Ushinomiya-cho, Sakyo-ku, Kyoto 606-8501, Japan}
\email{nakagawa.yujin@jamstec.go.jp, ebisawa.ken@jaxa.jp, enoto@kusastro.kyoto-u.ac.jp}


\maketitle

\begin{abstract}
Emission mechanism of the magnetars is still controversial while various
observational and theoretical studies have been made.
In order to investigate mechanisms of both the
persistent X-ray emission and the burst emission of the magnetars, we have proposed a 
model that the persistent X-ray emission
consists of numerous micro-bursts of various sizes.
If this model is correct, intensity Root Mean Square (RMS) variations
of the persistent emission
exceed the values expected from the Poisson distribution.
Using {\it Suzaku} archive data of  11 magnetars (22 observations),
the RMS intensity variations were calculated from 0.2\,keV to 70\,keV.
As a result, we found significant excess RMS intensity variations from
all the 11 magnetars.
We suppose that numerous mircro-bursts constituting the persistent X-ray emission
cause the observed variations, suggesting 
that the persistent X-ray emission and the burst emission have identical emission mechanisms. 
In addition, we found that the RMS intensity variations clearly increase
toward higher energy bands for 4 magnetars (6 observations).
The energy dependent RMS intensity variations imply that the
soft thermal component and the hard X-ray component
are emitted from different regions far apart from each other.
\end{abstract}

\section{Introduction}\label{sec:intro}
Magnetars are highly magnetized neutron stars \citep{duncan1992}, and
unique astrophysical objects to study physical phenomena under
extremely high magnetic field strengths greater than the quantum critical level 4.4$\times$10$^{13}$\,G (e.g., \cite{lyne2006}).
Among several classes of  magnetars, soft gamma repeaters (SGRs) and anomalous X-ray pulsars (AXPs)
are known to exhibit particularly
intriguing X-ray emitting phenomena.
While both exhibit persistent X-ray emission with typical
luminosities of $\sim10^{34}$--$10^{35}$\,erg\,s$^{-1}$ in 2--10\,keV,
the SGRs and some AXPs occasionally exhibit sporadic short bursts with typical durations of $\sim$100\,ms and
luminosities of $\sim10^{39}$--$10^{40}$\,erg\,s$^{-1}$ in 2--100\,keV.
These unusual phenomena are thought to be caused by
extremely strong magnetic field dissipation \citep{duncan1992}.

Models which reproduce spectra of the persistent X-ray emission of the magnetars were
studied based on observations by RXTE \citep{kuiper2004},
INTEGRAL \citep{molkov2005, rea2009} and {\it Suzaku}
\citep{esposito2007, nakagawa2009b, enoto2010a, enoto2010b, enoto2010c, 2017ApJEnoto}.
These observational studies suggest  that the magnetar persistent X-ray
spectra consist of a soft thermal ($<$10\,keV) component and a hard X-ray ($>$10\,keV) component.
The soft thermal component is reproduced by two blackbody functions
(2BB) with typical temperatures of $\sim$0.5\,keV and $\sim$1.4\,keV,
or a blackbody plus a power-law model (BB$+$PL)
with a typical temperature of $\sim$0.5\,keV and a typical power-law photon
index of $\sim$3 (e.g., \cite{nakagawa2009a}).
The hard X-ray component is well reproduced by a power-law model (PL) with a typical power-law
photon index of $\sim$1 (e.g., \cite{enoto2010c}).
The hard X-ray component should have a cutoff in high energy greater than $\sim200$\,keV (e.g., \cite{enoto2010b}),
otherwise the energy flux in the high energy goes to infinity.
However, no clear evidence of the cutoff has been found
(\cite{li2017} for upper limits of gamma-ray emission in 0.1--10\,GeV), 
while its hint has been  reported \citep{yasuda2015}.

Energy spectra of 50 short bursts from SGR\,1806$-$20 and 5 short bursts from SGR\,1900$+$14
with typical luminosities of $\sim10^{39}$--$10^{40}$\,erg\,s$^{-1}$ in 2--100\,keV
were observed by High Energy Transient Explorer 2 (HETE-2) with a wide energy band of 2--400\,keV,
and phenomenologically reproduced by the 2BB model \citep{nakagawa2007}
or the 2BB$+$PL model \citep{nakagawa2011b}.
In  recent studies based on {\it Suzaku} observations, energy spectra of weak bursts with
luminosities of $\sim10^{36}$--$10^{37}$\,erg\,s$^{-1}$ in 2--40\,keV
from SGR\,0501$+$4516 \citep{nakagawa2011a} and AXP\,1E\,1547.0$-$5408 \citep{enoto2012},
which have lower luminosities than the typical short bursts,
are reproduced with the hard X-ray component (PL) and the soft thermal component (2BB).
Thus the energy spectra of the persistent X-ray emission, the typical short bursts and the weak bursts are
likely to be reproduced by the same spectral model.

Several physical
models have been proposed to explain emission mechanisms of the short bursts and the persistent X-ray emission.
One of the ideas for the bursts is that the bursts are caused by heating of
the magnetic corona due to local magnetic reconnections \citep{lyutikov2003}.
The soft thermal component of the persistent X-ray emission is
explained, e.g., by the Surface Thermal Emission and Magnetospheric Scattering (STEMS) model \citep{guver2006}, 
the Resonant Cyclotron Scattering (RCS) model \citep{lyutikov2006}.
Meanwhile, the hard X-ray component is
explained by, e.g., thermal bremsstrahlung at the neutron star surface \citep{thompson2005, beloborodov2007},
Compton scattering in high magnetic fields \citep{baring2007, fernandez2007},
synchrotron emission in the magnetosphere \citep{heyl2005b},
a fallback disk model \citep{trumper2010} or a photon splitting model \citep{enoto2010c}.

Based on the unified analysis of the persistent X-ray emission and the burst emission,
two types of the correlation are reported: One is between the low and high temperatures of the 2BB
components (2BB temperature correlation; \cite{nakagawa2009a}).
The other is between the luminosities of the soft (2BB) and the hard (PL) components
over five orders of magnitude (luminosity correlation; \cite{nakagawa2011a}).

Based on the unified spectral analysis, the 2BB temperature correlation,
the luminosity correlation, and analogy with a relation between the solar microflare and the solar flare,
we have proposed a new idea named ''micro-burst model'' 
that the persistent X-ray emission is composed of numerous micro-bursts
of various sizes \citep{nakagawa2009a, nakagawa2011a}.
The micro-bursts may have a duration much less than $\sim$100\,ms,
a typical duration of the short bursts.
If the persistent X-ray emission is composed of such numerous micro-bursts,
a cumulative number-intensity distribution of the micro-bursts would show a power-law
distribution which has been found for typical short bursts (e.g., \cite{nakagawa2007}).
Such power-law distribution is often  referred to as the Gutenberg-Richter law
\citep{gutenberg1956}.

We have calculated expected fluxes of the putative micro-bursts constituting the persistent X-ray emission by extrapolating the
cumulative number-intensity distribution of typical bursts
observed by HETE-2 for SGR\,1806$-$20 \citep{nakagawa2011b}.
We found that the expected flux, accumulating the unresolved micro-bursts,
is comparable to the observed persistent X-ray fluxes.
A similar study was performed on an outburst of  AXP\,1E\,1547.0$-$5408 with {\it Suzaku} \citep{enoto2012}.

If the persistent X-ray emission is not static, but composed of numerous micro-bursts of various sizes
following a particular cumulative number-intensity distribution,
dispersion of the micro-burst intensities, as well as the persistent X-ray flux,
should exceed the value expected from the Poisson distribution.
In order to measure the dispersion quantitatively, in this paper, we
calculate Root Mean Square (RMS) intensity variations in the persistent
X-ray emission  using the {\it Suzaku} data.
{\it Suzaku} has great capabilities to estimate the RMS intensity variations,
because the on-board narrow field instruments of X-ray imaging spectrometer (XIS; 0.2--12\,keV; \cite{koyama2007})
and the hard X-ray detector (HXD; 10--700\,keV; \cite{takahashi2007}) have high sensitivities and wide energy bands.

\section{Observation and Data Reduction}
The present studies are performed using
 {\it Suzaku} archive data of 11 magnetars (22 observations).
Table \ref{tab:suzaku_obs_list} shows a summary of the observations.
Besides, {\it Suzaku} archive data include the magnetars
AXP\,Swift\,J1834.9$-$0846 and AXP\,1E\,1841$-$045, which
are not used in the present study due to no significant detection and
contamination by a nearby supernova remnant, respectively.
Since there are not enough photon counts in the
HXD-GSO energy band, we focus on the XIS and HXD-PIN data.
Although  HXD-PIN event data have  a time resolution of 61\, $\mu$s,  there are not enough photon counts
 to find any direct evidence of micro-bursts.

Reduction of the XIS and HXD-PIN event data were made using HEAsoft\,6.16 software.
The latest calibration database (CALDB:\,20150312) was applied to unfiltered XIS event data using {\it aepipeline} (v1.1.0).
We created light curves and spectra from the cleaned XIS event data using {\it xselect} (v2.4c).
Response matrix files were generated by {\it xisrmfgen} (v2012-04-21), and ancillary response function files by {\it xissimarfgen} (v2010-11-05).
The net exposures of the XIS data are summarized in table \ref{tab:suzaku_obs_list}.

The latest calibration database (CALDB:\,20110915) was applied to unfiltered HXD-PIN event data using {\it aepipeline} (v1.1.0).
We created light curves and spectra using {\it xselect}.
Dead time corrections were applied to the spectra using {\it hxddtcor} (v1.50)
as well as to the light curves using the recipe published on a {\it Suzaku}
website\footnote{http://www.astro.isas.jaxa.jp/suzaku/analysis/hxd/hxdfaq/hxd\_dtcor\_lc.html (last accessed on 2017-10-16).}.
Response matrix files were taken from
CALDB\footnote{We choose the response matrix files for each observation according to
the instruction at http://www.astro.isas.jaxa.jp/suzaku/analysis/hxd/pinnxb/quick/index.html.}.
The net exposures of the HXD-PIN data are summarized in table \ref{tab:suzaku_obs_list}.

\section{Data Analysis}\label{sec:data_analysis}
For each observation, the  XIS events  were extracted from box regions centered on the objects,
while the background events  were extracted from box regions around the objects
with the same  area.
The HXD-PIN background was subtracted using the background files supplied by Suzaku Guest Observer 
Facility\footnote{https://heasarc.gsfc.nasa.gov/docs/suzaku/analysis/pinbgd.html (last accessed Oct 16, 2017)}.
The quoted errors hereafter refer to 68\% confidence levels.

In this paper, we define the RMS intensity variations $R$ as
\begin{equation}\label{rms_eq}
R = \frac{
\bigl[
\frac{1}{N-1}
\bigl\{
\sum_{i}(x_i - \overline{x})^2 - \sum_{i}{\delta_{x_i}}^2
\bigr\}
\bigr]^{\frac{1}{2}}
}{\overline{x}},
\end{equation}
where $i$ is the bin number, $x_i$ is the  background-subtracted counts
per bin, $\overline{x}$ is the
average of $x_i$, $\delta_{x_i}$ is the error of $x_i$ and $N$ is the number of bins.
Here, $N$  is obtained by dividing the net exposures in table \ref{tab:suzaku_obs_list} by the time bin-widths.

We calculated RMS intensity variations for each observation using the background-subtracted light curves
in the 0.2--12\,keV (XIS) and 10--70\,keV (HXD-PIN) energy bands.
Time resolutions of the light curves are 8\,s in 0.2--12\,keV and 128\,s in 10--70\,keV.
The RMS intensity variations are found to be $R_{\mathrm{X}}=$1.3--135\% in the 0.2--12\,keV energy band
and $R_{\mathrm{P}}=$17--99\% in the 10--70\,keV energy band, depending on sources and observations (table \ref{tab:suzaku_obs_list}).

Based on Monte Carlo simulations, we confirmed that variations due to rotations
of the magnetars ($\sim$2--12\,s; e.g., \cite{enoto2010c}),
and long-term ($\sim$1\,day) flux variations do not affect the RMS intensity variations
when using the 8\,s (0.2--12\,keV) light curves.
We also estimated variations caused by background fluctuations  using {\it Suzaku} data of 
hard and bright non-variable sources (table \ref{tab:suzaku_obs_list_bkg}),
and confirmed that the background fluctuations are not significant for most cases.

Next, we estimate effects of obviously  bright bursts, which  would significantly affect  RMS intensity variations.
We performed burst search using the 0.2--12\,keV light curves
with a 8\,s time resolution of the XIS.
We searched for such bright bursts in the light curves that exceed $\lambda+5\sigma$, where $\lambda$ is 
the average and $\sigma$ is the standard deviation.
Then we identified visually obvious bright bursts from 7 out of 22 observations   as summarized in table \ref{tab:suzaku_obs_list}.
After removing the bright bursts from the light curves, we calculated the RMS variations as before.
Consequently, the RMS intensity variations without influence
of the bright burst emission are  $R'_{\mathrm{X}}=$1.3--18.8\% in the 0.2--12\,keV energy band
and $R'_{\mathrm{P}}=$17--99\% in the 10--70\,keV energy band as summarized in table \ref{tab:suzaku_obs_list}.
These variations are considered to be intrinsic variations of the persistent emission.

We also calculated the RMS intensity variations with finer energy bands
using the background-subtracted light curves of the XIS with a time resolution of 32\,s.
Figure \ref{fig:rms_spc_summary} shows energy dependency of the RMS intensity variations
with $E^2{f}(E)$ spectra where $f(E)$ is the photon spectrum for each observation.
The $E^2{f}(E)$ spectra are  consistent with \citet{2017ApJEnoto}.
Among the observations, the RMS intensity variations clearly increase toward higher energy bands for 4 magnetars (6 observations)
as shown in the panels (a), (b), (d), (h), (m) and (o) in figure \ref{fig:rms_spc_summary}.

\section{Result}
\subsection{Summary of Data Analysis}
We calculated RMS intensity variations using the {\it Suzaku} data archive for 11 magnetars (22 observations).
The RMS intensity variations are significantly greater than the values expected from the Poisson distribution
for all the 22  observations of 11 magnetars in the 0.2--12\,keV energy band (XIS) and 5 magnetars
in the 10--70\,keV energy band (HXD-PIN).
For these 5 magnetars, there were 15 observations, and significant variation was detected from 8 out of them.

\subsection{Mathematical Formulation of RMS Intensity Variations}\label{sec:math_rms}
In order to understand  observed  RMS intensity variations,
we calculate expected RMS intensity variations with mathematical approach.
We define the expected RMS intensity variations as
\begin{equation}\label{eq:rms_model}
R_{\mathrm{M}} = (\sigma_{\mathrm{c}}^2 - \sigma_{\mathrm{p}}^2)^{\frac{1}{2}}S_{\mathrm{a}}^{-1},
\end{equation}
where $\sigma_{\mathrm{c}}^2$ is variance of the expected cumulative number-intensity distribution,
$\sigma_{\mathrm{p}}^2$ is variance of  Poisson distribution,
and $S_{\mathrm{a}}$ is
 an average fluence of the micro-bursts.
We assume that $\sigma_{\mathrm{c}}^2$, $\sigma_{\mathrm{p}}^2$ and $S_{\mathrm{a}}$
are defined as values  in the 0.2--12\,keV energy band.
The expected RMS intensity variations are independent of the observation exposure time
exceeding 1\,ms, if we assume each micro-burst has 1\,ms duration.

\begin{figure}
  \begin{center}
     \begin{minipage}{0.48\hsize}
	\includegraphics[angle=90,height=7.19cm]{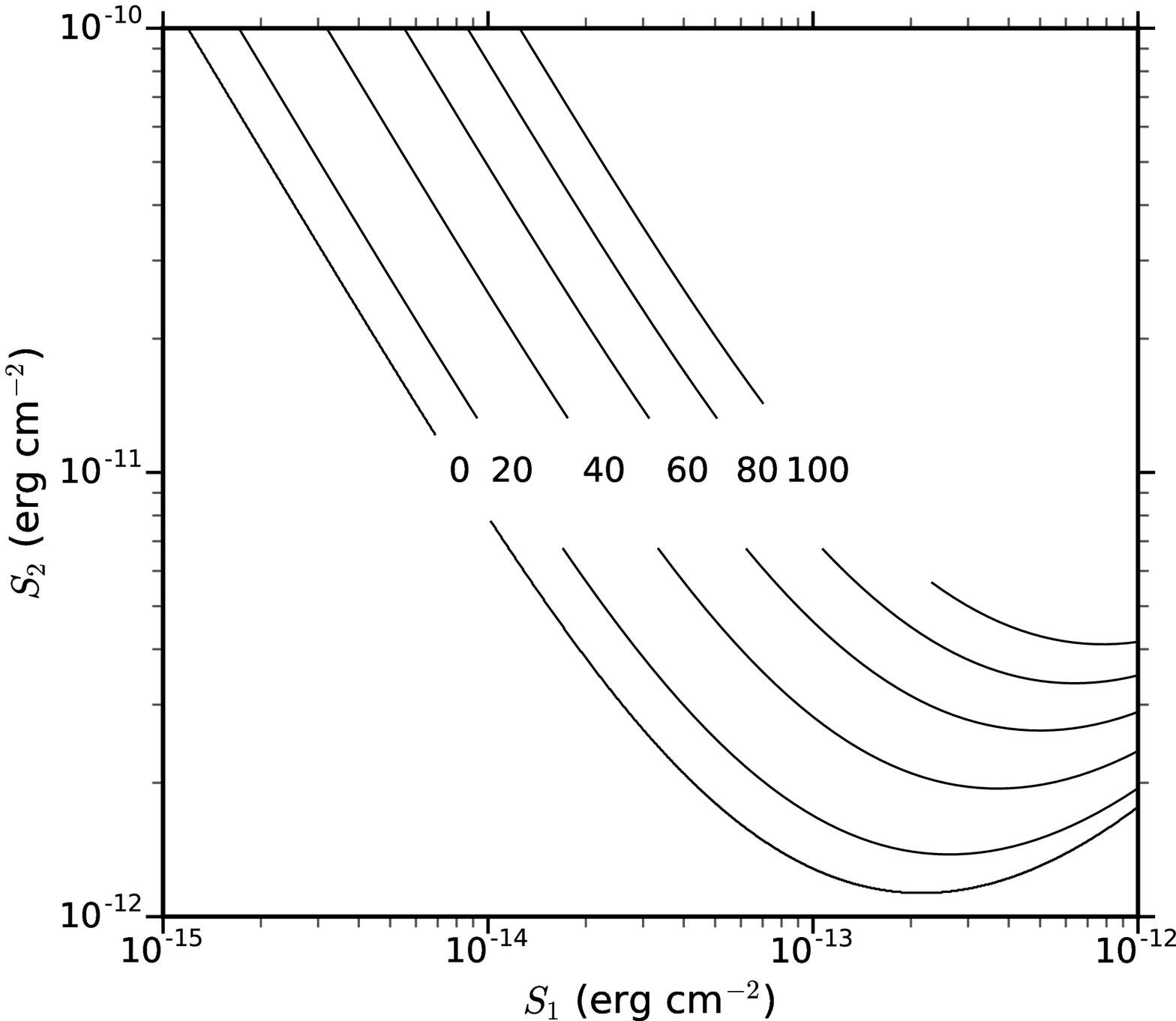}
     \end{minipage}
     \begin{minipage}{0.48\hsize}
	 \includegraphics[angle=90,height=7.19cm]{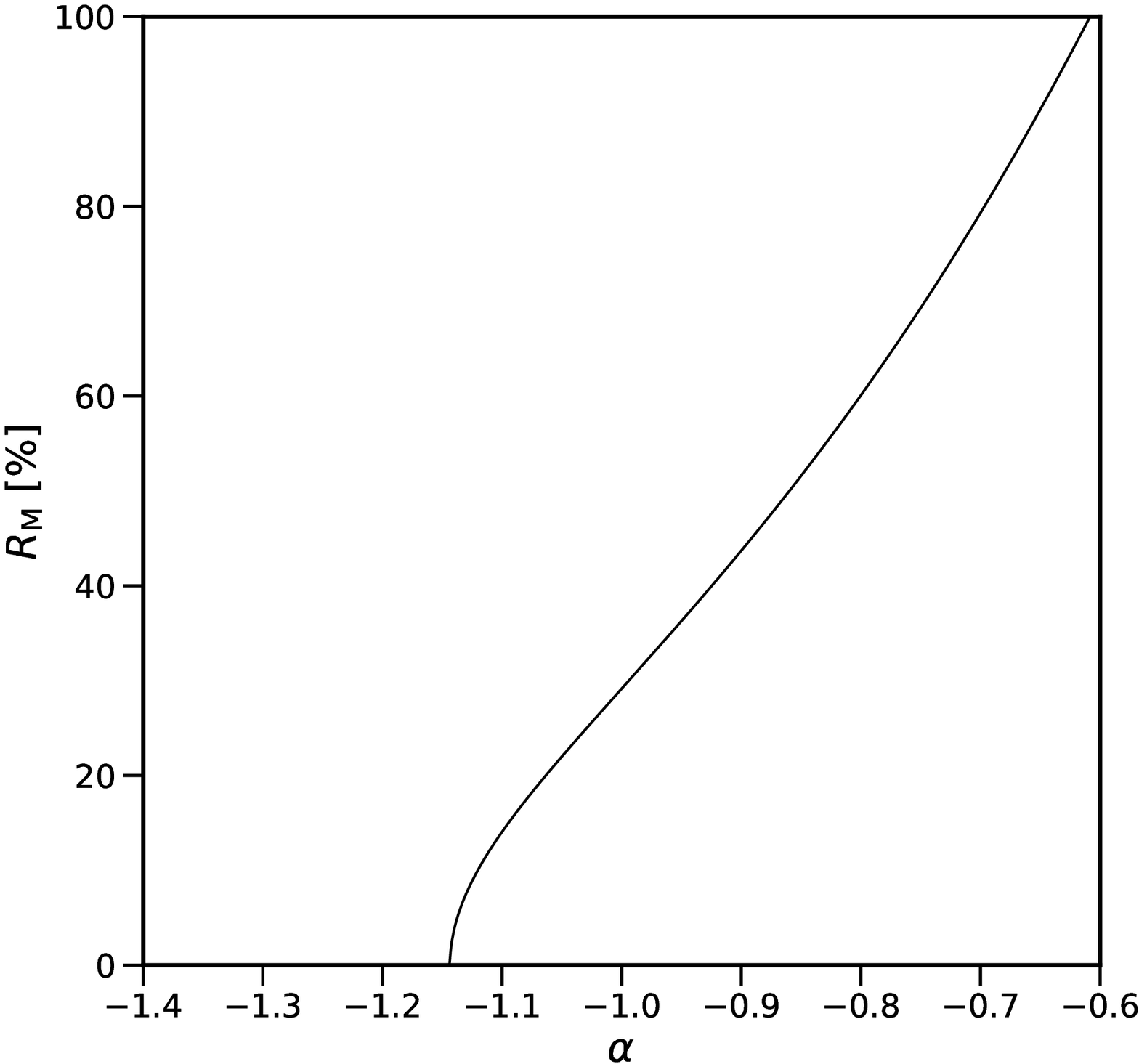}
     \end{minipage}
  \end{center}
  \caption{
  Left: A two-dimentional contour graph of $R_{\mathrm{M}}$ with respect to $S_1$ and $S_2$.
  Contour lines are plotted for $R_{\mathrm{M}}$ ranging from 0 to 100.
  Right: A relation between $\alpha$ and $R_{\mathrm{M}}$.
  \label{fig:logn_logs_map}
  }
  \label{fig:relation_alpha_rms}
\end{figure}

The expected cumulative number-intensity distribution from a single magnetar is defined as
\begin{equation}\label{eq:logn_logs}
N_{\mathrm c}(>S_{\mathrm c}) = A_{\mathrm c}S_{\mathrm c}^{\alpha},
\end{equation}
where $A_{\mathrm c}$ is a normalization, $S_{\mathrm c}$ is a fluence
of micro-bursts, $\alpha$ is an index,
and $N_{\mathrm c}(>S_{\mathrm c})$ is
a cumulative number of micro-bursts whose fluences are greater than $S_{\mathrm c}$.
The expected cumulative number-intensity distribution for a hypothetical magnetar
is assumed to have $\alpha =-1.1$ \citep{nakagawa2007} 
and $A_{\mathrm c} = 7\times$10$^{-9}$\,bursts\,day$^{-1}$ at $S_{\mathrm c} = 1$\,erg\,cm$^{-2}$.
A probability density function of the expected cumulative number-intensity distribution
is defined as
\begin{eqnarray}\label{eq:pdf}
P(S_{\mathrm{c}}) &=& N'_{\mathrm c}(>S_{\mathrm c}) \left( \int_{S_1}^{S_2} N'_{\mathrm c}(>S_{\mathrm c}) dS_{\mathrm c} \nonumber \right)^{-1} \\
&=& \frac{{\alpha}S_{\mathrm c}^{\alpha-1}}{S_2^{\alpha} - S_1^{\alpha}},
\end{eqnarray}
where $S_1$ and $S_2$ ($S_1 < S_2$) are minimum and maximum fluences of the interval of $S_{\mathrm c}$
which satisfy the probability density function.
Using  equation (\ref{eq:pdf}), variance of the expected cumulative number-intensity distribution is calculated  as
\begin{eqnarray}\label{eq:logn_logs_vari}
\sigma_{\mathrm{c}}^2 &=& E[S_{\mathrm c}^2] - (E[S_{\mathrm c}])^2 \nonumber \\
&=& \int_{S_1}^{S_2} S_{\mathrm c}^2 P(S_{\mathrm{c}}) dS_{\mathrm c}
- \left\{ \int_{S_1}^{S_2} S_{\mathrm c} P(S_{\mathrm{c}}) dS_{\mathrm c} \right\}^{2} \nonumber \\
&=&
-\alpha{A_{\mathrm c}}f_{\mathrm m}^{-1}
\int_{S_1}^{S_2} S_{\mathrm c}^{\alpha+1} dS_{\mathrm c}
-\left\{{\alpha{A_{\mathrm c}}}{f_{\mathrm m}^{-1}}
\int_{S_1}^{S_2} S_{\mathrm c}^{\alpha} dS_{\mathrm c} \right\}^{2}
\nonumber \\
&=&
\left\{ \begin{array}{ll}
\frac{\alpha}{\alpha+2}
{A_{\mathrm c}}f_{\mathrm m}^{-1}
\left(
S_1^{\alpha+2} - S_2^{\alpha+2}
\right)
-
\left\{
\frac{\alpha}{\alpha+1}
{A_{\mathrm c}}f_{\mathrm m}^{-1}\left(
S_1^{\alpha+1} - S_2^{\alpha+1}
\right)
\right\}^{2}
& (\alpha \ne -1 \land 0>\alpha>-2) \\
{A_{\mathrm c}}{f_{\mathrm m}^{-1}}\left(
S_1 - S_2
\right)
-
\left\{
{A_{\mathrm c}}{f_{\mathrm m}^{-1}}
\left(\log|S_1| - \log|S_2|\right)
\right\}^{2}
& (\alpha = -1)
\end{array} \right.,
 \label{eq5}
\end{eqnarray}
where $E[S_{\mathrm c}^2]$ is an expectation of $S_{\mathrm c}^2$,
$E[S_{\mathrm c}]$ is an expectation of $S_{\mathrm c}$ and
$f_{\mathrm m} = -A_{\mathrm c}(S_2^\alpha - S_1^\alpha)$ is a frequency
of the micro-bursts with the fluences between $S_1$ and $S_2$.
Variance of the expected cumulative number-intensity distribution $\sigma_{\mathrm{c}}^2$
 depends on the frequency of micro-bursts $f_{\mathrm m}$ as well as the  fluence distribution index $\alpha$.
Hence, the values of $R_{\mathrm{M}}$ depend on choices of $S_1$, $S_2$ and $\alpha$.
Figure \ref{fig:logn_logs_map} (left) shows a two-dimentional contour graph of $R_{\mathrm{M}}$
with respect to $S_1$ and $S_2$, and figure \ref{fig:relation_alpha_rms} (right) shows a relation between $\alpha$ and $R_{\mathrm{M}}$;
other  parameter values are fixed to those determined below.

Short bursts are known to have spiky structures in 0.5\,ms light curves, 
which may be caused by a rapid  energy reinjection and cooling \citep{nakagawa2007}.
We assume 1\,ms durations for the micro-bursts 
under the assumption that one cycle of the energy reinjection and
cooling corresponds to a single  micro-burst.
Using the assumed duration of 1\,ms for the micro-bursts and typical flux of $10^{-11}$\,erg\,cm$^{-2}$\,s$^{-1}$
for the persistent X-ray emission, we assume $S_1 = 0.001$\,s\,$ \times $\,$10^{-11}$\,erg\,cm$^{-2}$\,s$^{-1} = 10^{-14}$\,erg\,cm$^{-2}$.
We also assume $S_2 = 10^{-11}$\,erg\,cm$^{-2}$ which is substantially below
a burst detection limit of {\it Suzaku} ($5\sigma \approx 10^{-10}$\,erg\,cm$^{-2}$ described in section \ref{sec:data_analysis}).
We found $\sigma_{\mathrm{c}}^2 = 5.8\times10^{-26}$\,erg$^2$\,cm$^{-4}$ from equation (\ref{eq5}).

The expected cumulative Poisson distribution is defined as
$N_{\mathrm p}(>k') = A_{\mathrm p}(1-\mathrm{e}^{-\lambda}\sum^{k'}_{k=0}\lambda^{k}/k!)$,
where $A_{\mathrm p}$ is a normalization,
$\lambda$ is mean counts per bin of a light curve,
$k'$ and $k$ are integer values, and $N_{\mathrm p}(>k')$ is
a cumulative number of the bins corresponding to micro-bursts for which $k$ is greater than $k'$.
The expected cumulative Poisson distribution for the hypothetical magnetar 
has $\lambda = 12.12$\,counts\,(2s)$^{-1}$
in the 0.2--12\,keV energy band \citep{nakagawa2011a} and an assumed normalization of
$A_{\mathrm p} = 2\times$10$^{5}$\,bursts\,day$^{-1}$.
We assume that $\lambda = 12.12$\,counts\,(2s)$^{-1}$  correspond to
an average fluence of the micro-bursts in the 0.2--12\,keV energy band of
$S_{\mathrm{a}} = 7.54\times$10$^{-13}$\,erg\,cm$^{-2}$.
We also assume that $k'$ has a fluence $S_{\mathrm{p}}(k')$ where
$S_{\mathrm{p}}(k'+1) - S_{\mathrm{p}}(k') = S_{\mathrm{p}}(1) = \lambda^{-1}S_{\mathrm{a}}$ is 6.22$\times$10$^{-14}$\,erg\,cm$^{-2}$.
We found the variance of the expected Poisson distribution as 
$\sigma_{\mathrm{p}}^2 = {\lambda^{-1}}S_{\mathrm{a}}^2 = 4.7\times10^{-26}$\,erg$^2$\,cm$^{-4}$.

Finally, we obtain $R_{\mathrm{M}} = 14$\% for the hypothetical magnetar
by substituting $\sigma_{\mathrm{c}}^2$, $\sigma_{\mathrm{p}}^2$ and $S_{\mathrm{a}}$
shown above in  equation (\ref{eq:rms_model}). 
This is consistent with the observed values of $R'_{\mathrm{X}} = $1.3--18.8\%.
Difference of the RMS intensity variations among magnetars may be explained by
difference of the fluence distribution index $\alpha$.

\subsection{Comparison of  cumulative number-intensity distributions}
\begin{figure}
  \begin{center}
     \includegraphics[angle=90,width=8.19cm]{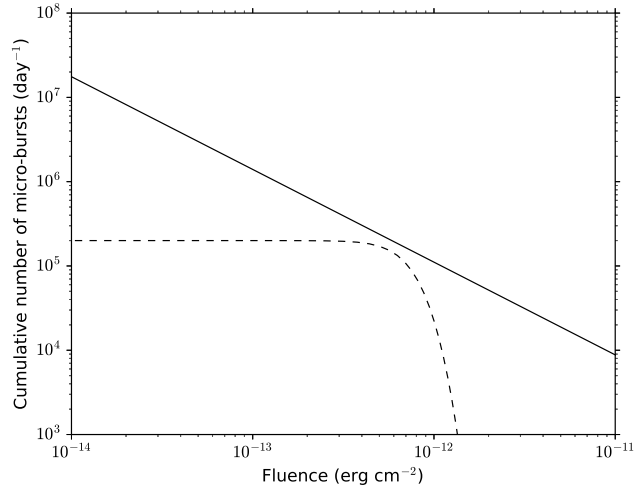}
  \end{center}
  \caption{
  An expected cumulative number-intensity distribution of the micro-bursts (solid line)
  and an  assumed cumulative Poisson distribution (dashed line).
  See details in text.
  }
  \label{fig:comp_poisson_cumulative_ns}
\end{figure}

Figure \ref{fig:comp_poisson_cumulative_ns} shows  comparison between the 
 expected cumulative number-intensity distribution of the micro-bursts and that 
expected from Poisson distribution for a  hypothetical magnetar. 
We see that the former 
clearly has a wider  distribution
than the latter.
In particular, 
reduction  of the bursts  above $\sim2\times10^{-12}$\,erg\,cm$^{-2}$
makes the expected cumulative number-intensity distribution closer to
that expected from Poisson distribution. 
We confirmed that the RMS intensity variations significantly exceed the values expected from the Poisson distribution
even after removing bright bursts (section \ref{sec:data_analysis}).
The wider cumulative number-intensity distribution can 
naturally explain  the observed excess RMS intensity variations.
Thus, the observation is consistent with the assumption 
that the persistent X-ray emission is composed of
numerous mircro-bursts of various sizes subject to a particular cumulative number-intensity distribution.

\section{Discussion}\label{sec:discussion}
\subsection{Expected Flux from Cumulative Number-Intensity Distribution}
Integrating energies of the micro-bursts from  $S_{\mathrm c} =
 10^{-14}$\,erg\,cm$^{-2}$ to 10$^{-11}$\,erg\,cm$^{-2}$
using the expected cumulative number-intensity distribution
in section \ref{sec:math_rms} \citep{nakagawa2007},
we obtain the persistent X-ray flux of a hypothetical magnetar as 
$\sim$1.1$\times$10$^{-11}$\,erg\,cm$^{-2}$\,s$^{-1}$. This
is comparable with  a typical observed flux of $\sim$9.9$\times$10$^{-12}$\,erg\,cm$^{-2}$\,s$^{-1}$
for SGR\,1806$-$20 \citep{nakagawa2009b}.
This result is consistent with the persistent X-ray  fluxes 
estimated from observed cumulative number-intensity distributions
obtained by \citet{nakagawa2011b} and \citet{enoto2012}.

\subsection{Energy Dependent RMS Intensity Variations}
\begin{figure*}
   \begin{center}
   \begin{minipage}{0.32\hsize}
      \includegraphics[height=2.7cm,width=5.46cm]{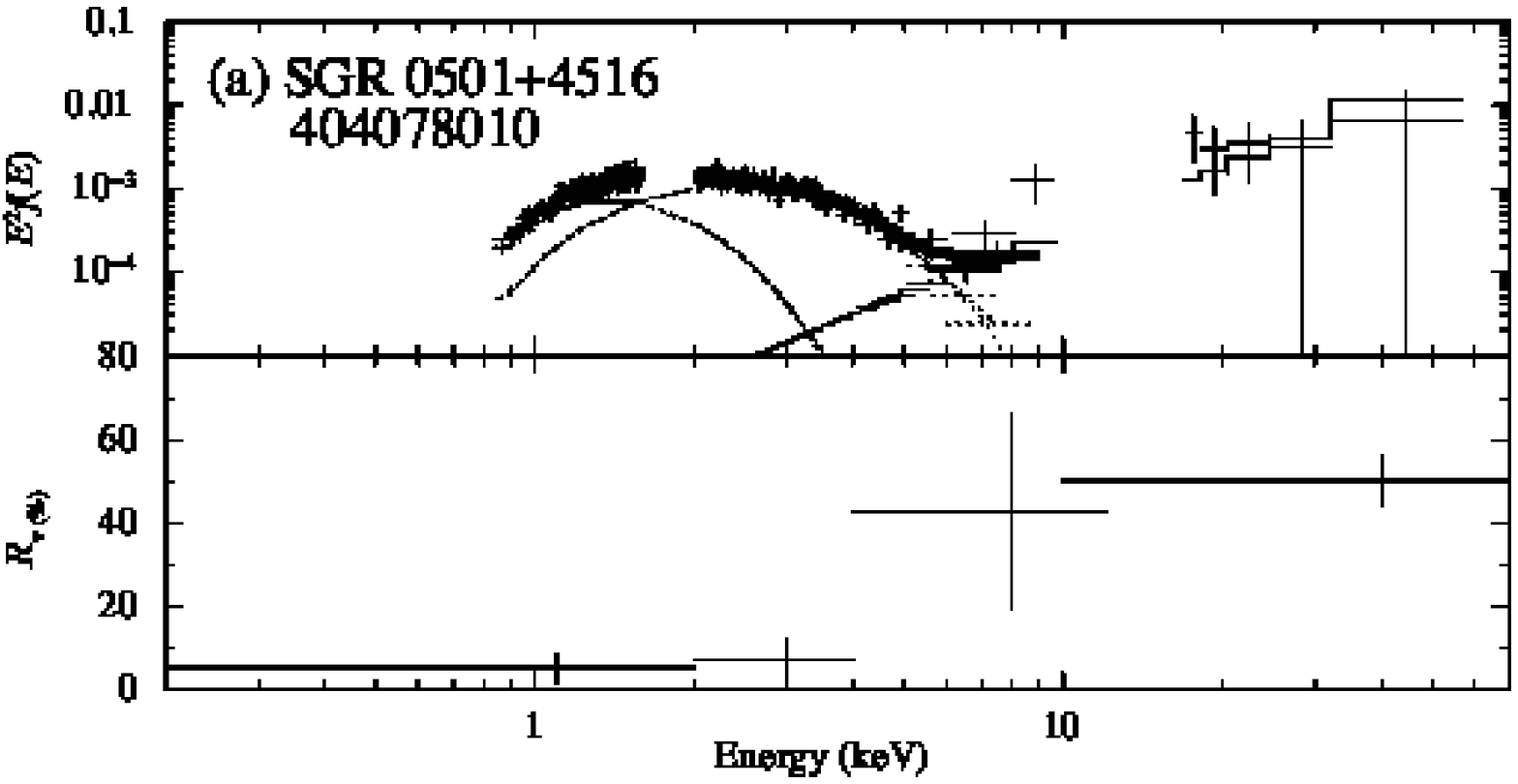}
   \end{minipage}
   \begin{minipage}{0.32\hsize}
     \includegraphics[height=2.7cm,width=5.46cm]{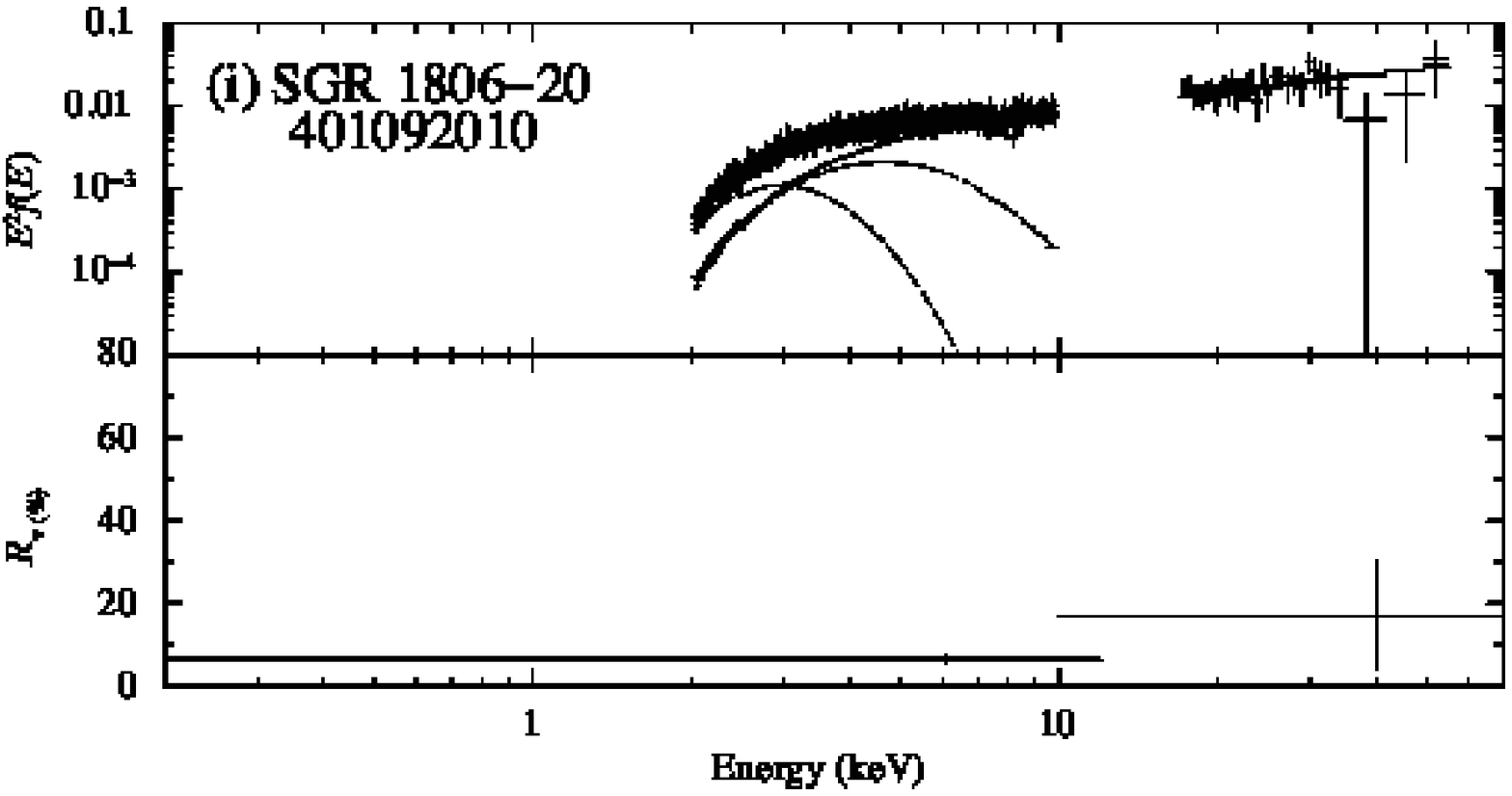}
   \end{minipage}
   \begin{minipage}{0.32\hsize}
     \includegraphics[height=2.7cm,width=5.46cm]{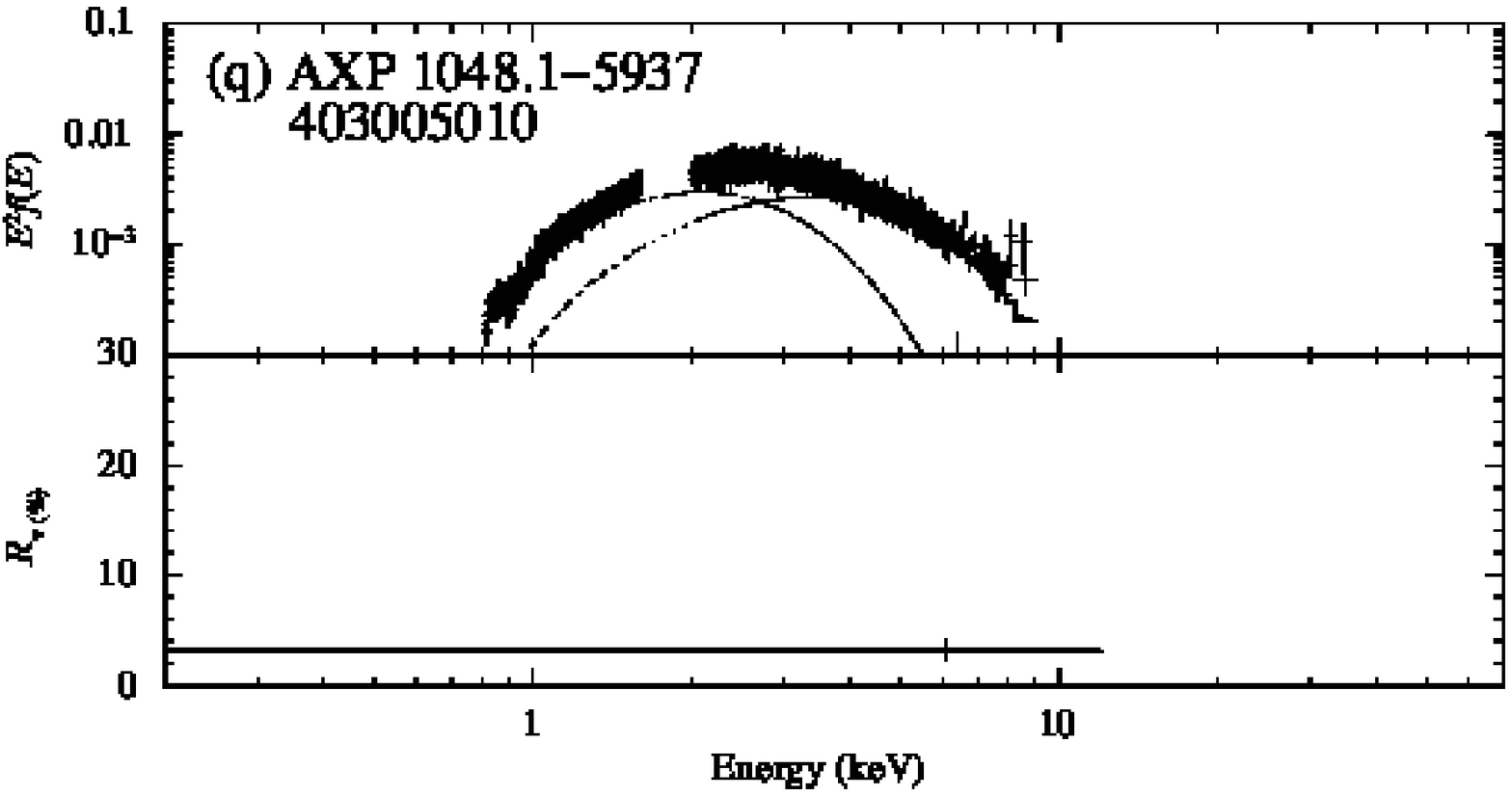}
   \end{minipage}
   \\
   \begin{minipage}{0.32\hsize}
     \includegraphics[height=2.7cm,width=5.46cm]{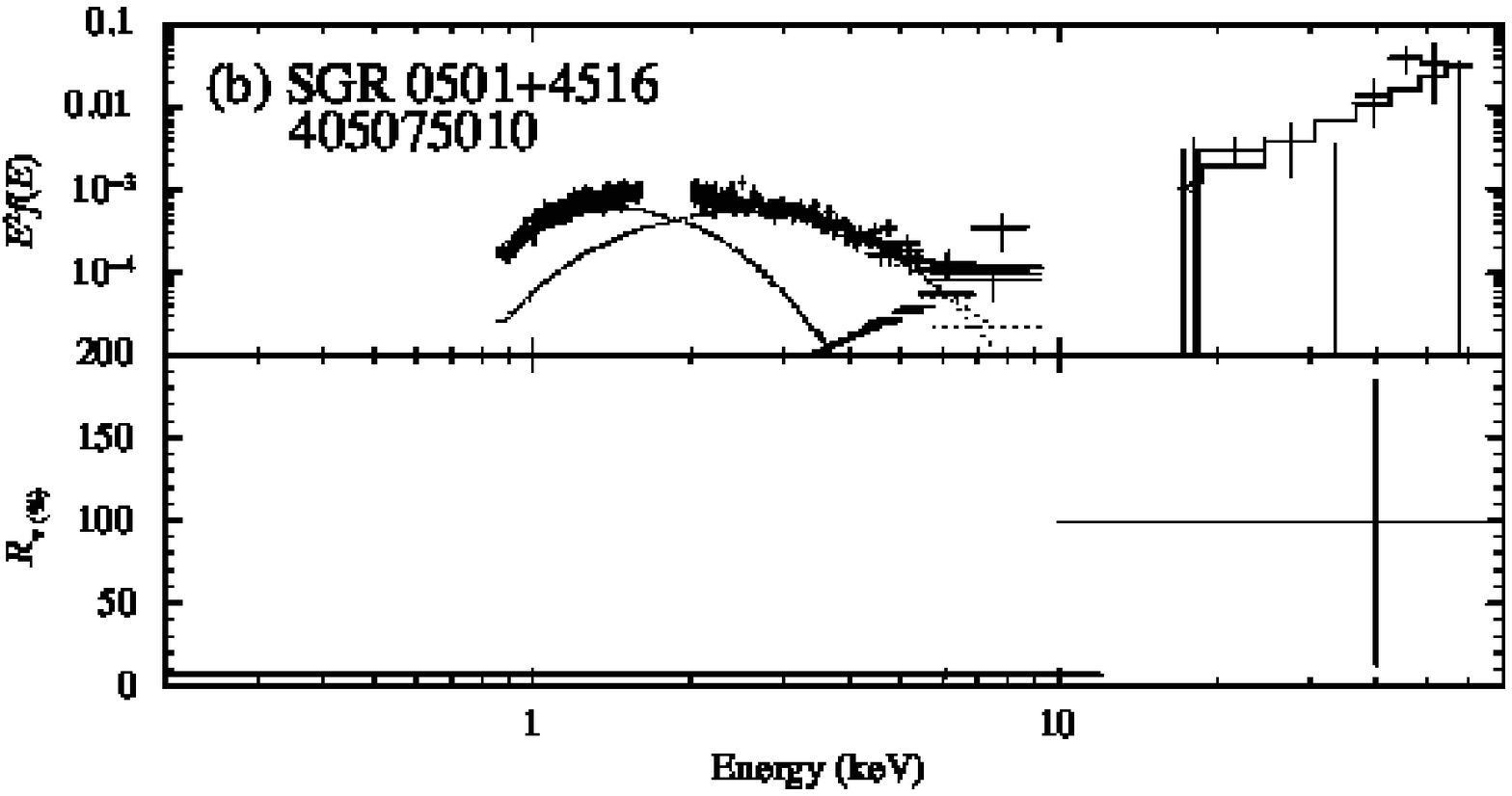}
   \end{minipage}
   \begin{minipage}{0.32\hsize}
      \includegraphics[height=2.7cm,width=5.46cm]{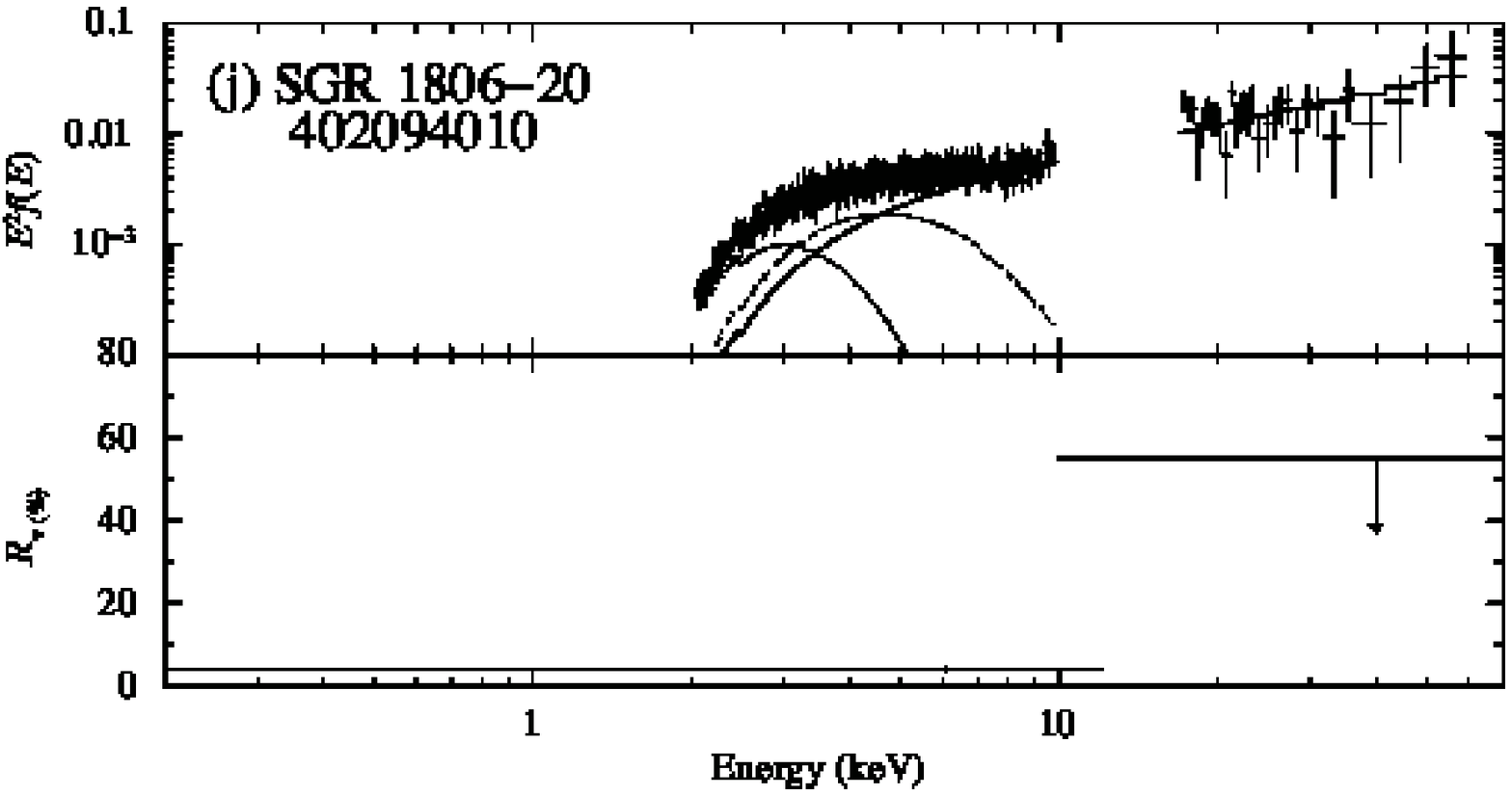}
   \end{minipage}
   \begin{minipage}{0.32\hsize}
      \includegraphics[height=2.7cm,width=5.46cm]{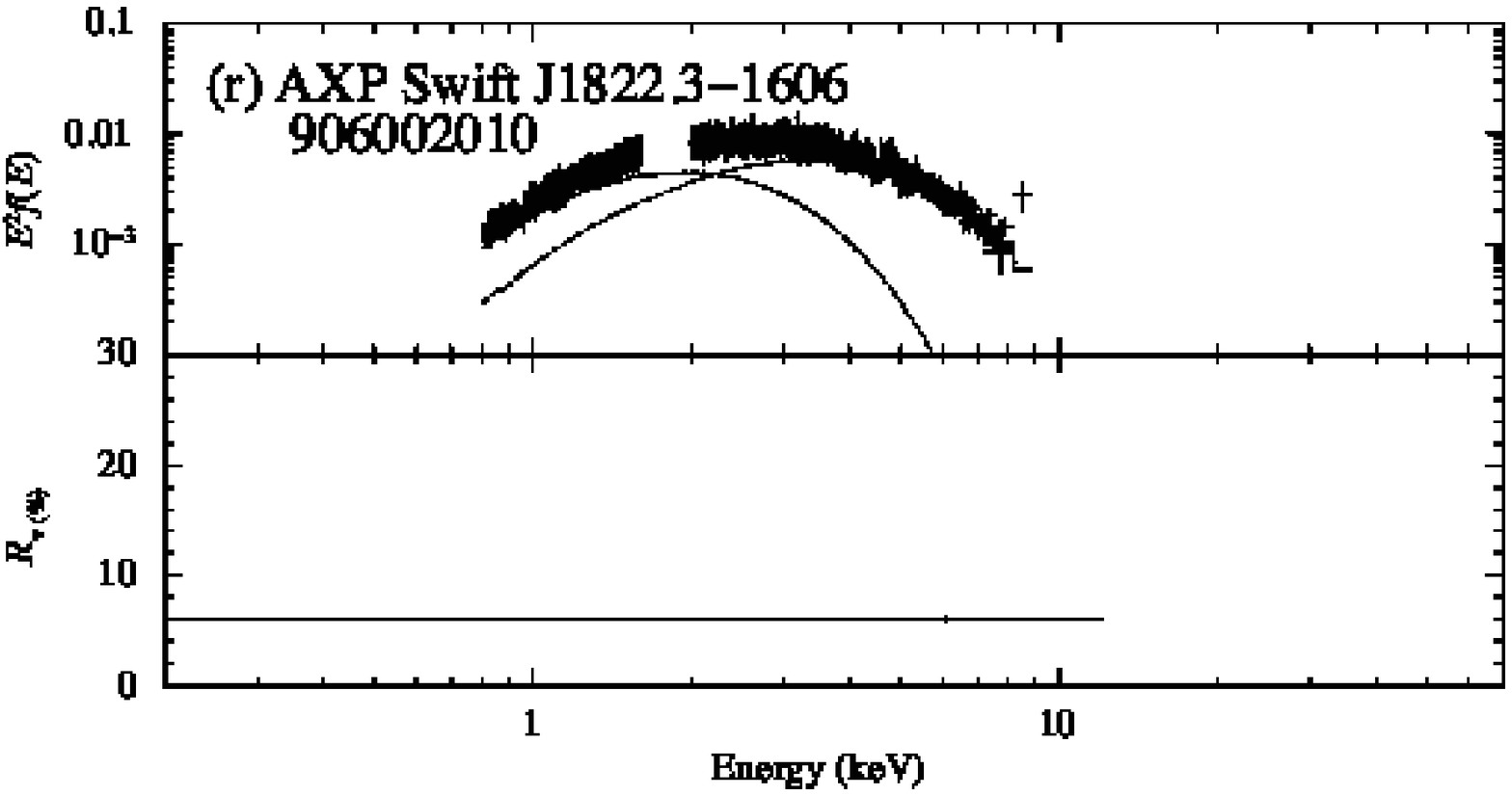}
   \end{minipage}
   \\
   \begin{minipage}{0.32\hsize}
 	\includegraphics[height=2.7cm,width=5.46cm]{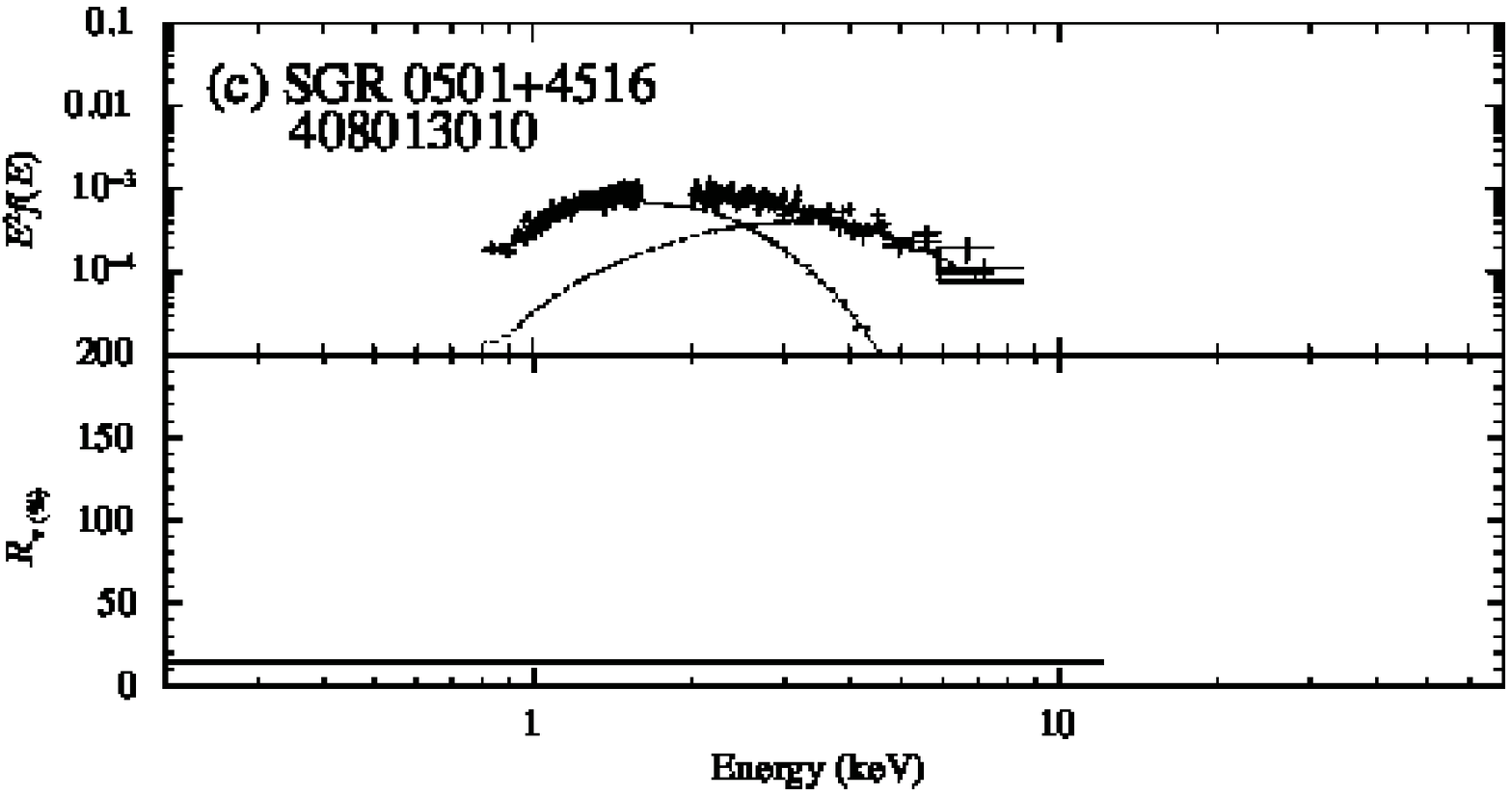}
   \end{minipage}
   \begin{minipage}{0.32\hsize}
     \includegraphics[height=2.7cm,width=5.46cm]{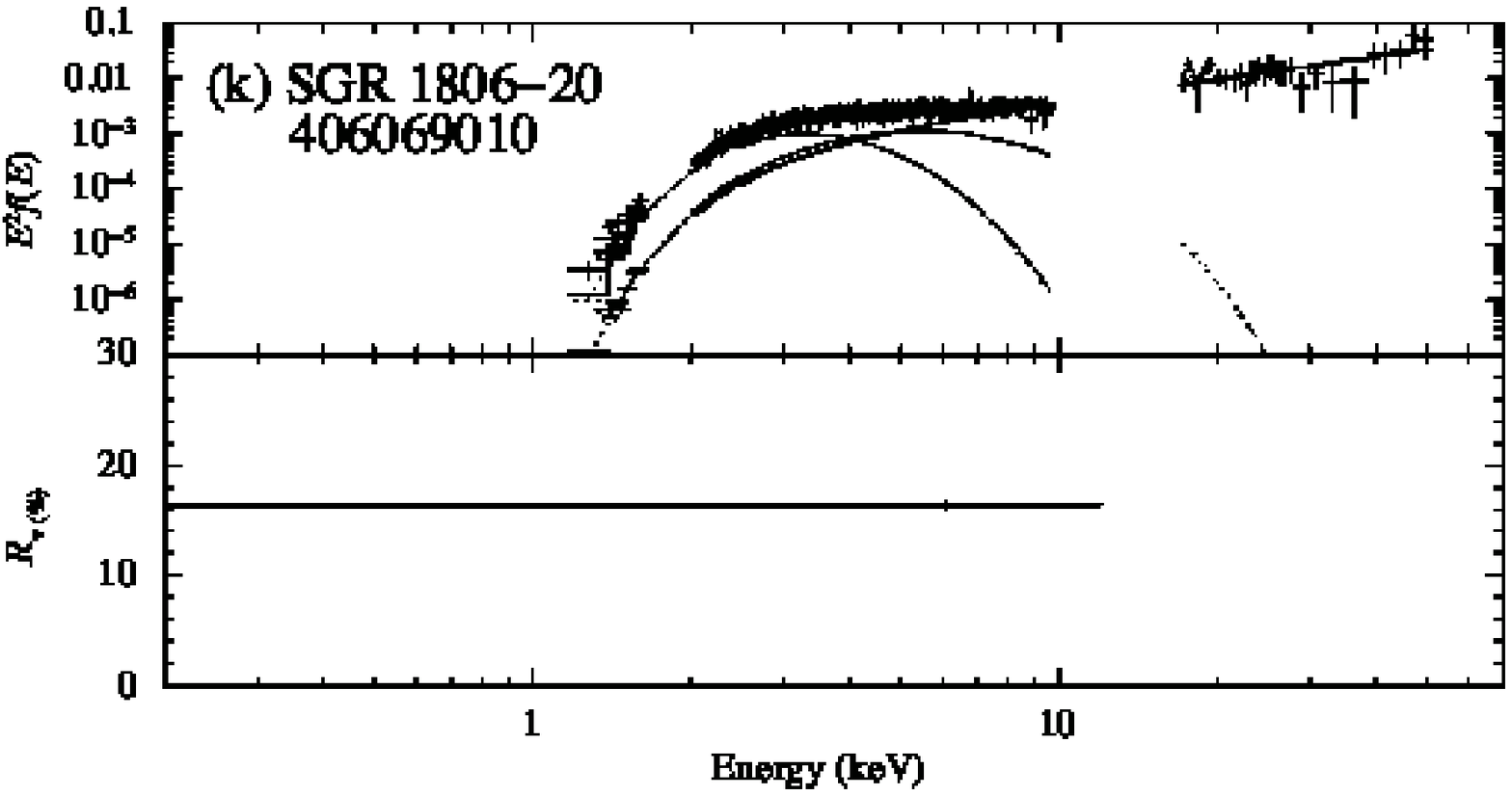}
   \end{minipage}
   \begin{minipage}{0.32\hsize}
     \includegraphics[height=2.7cm,width=5.46cm]{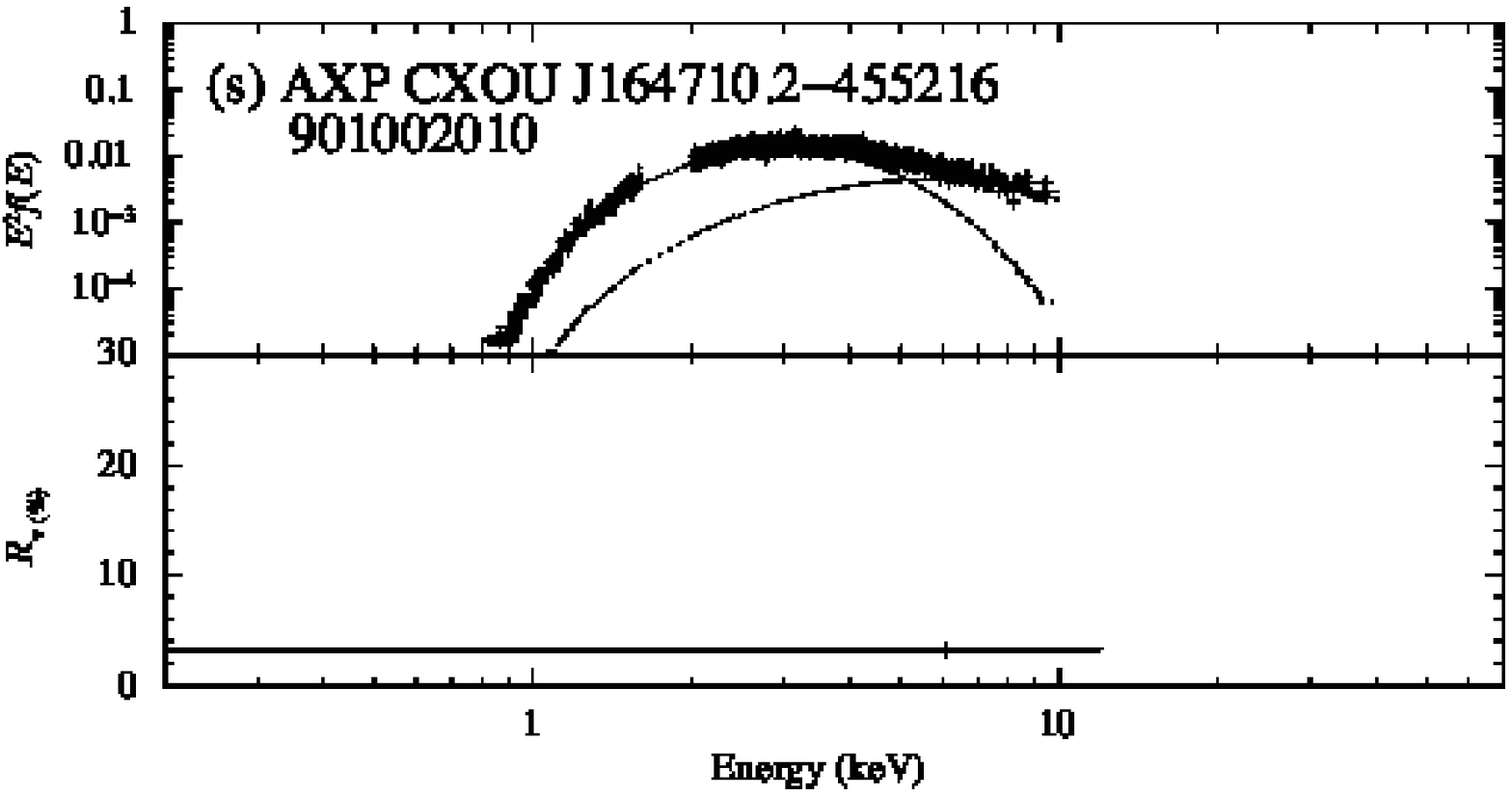}
   \end{minipage}
   \\
   \begin{minipage}{0.32\hsize}
   \includegraphics[height=2.7cm,width=5.46cm]{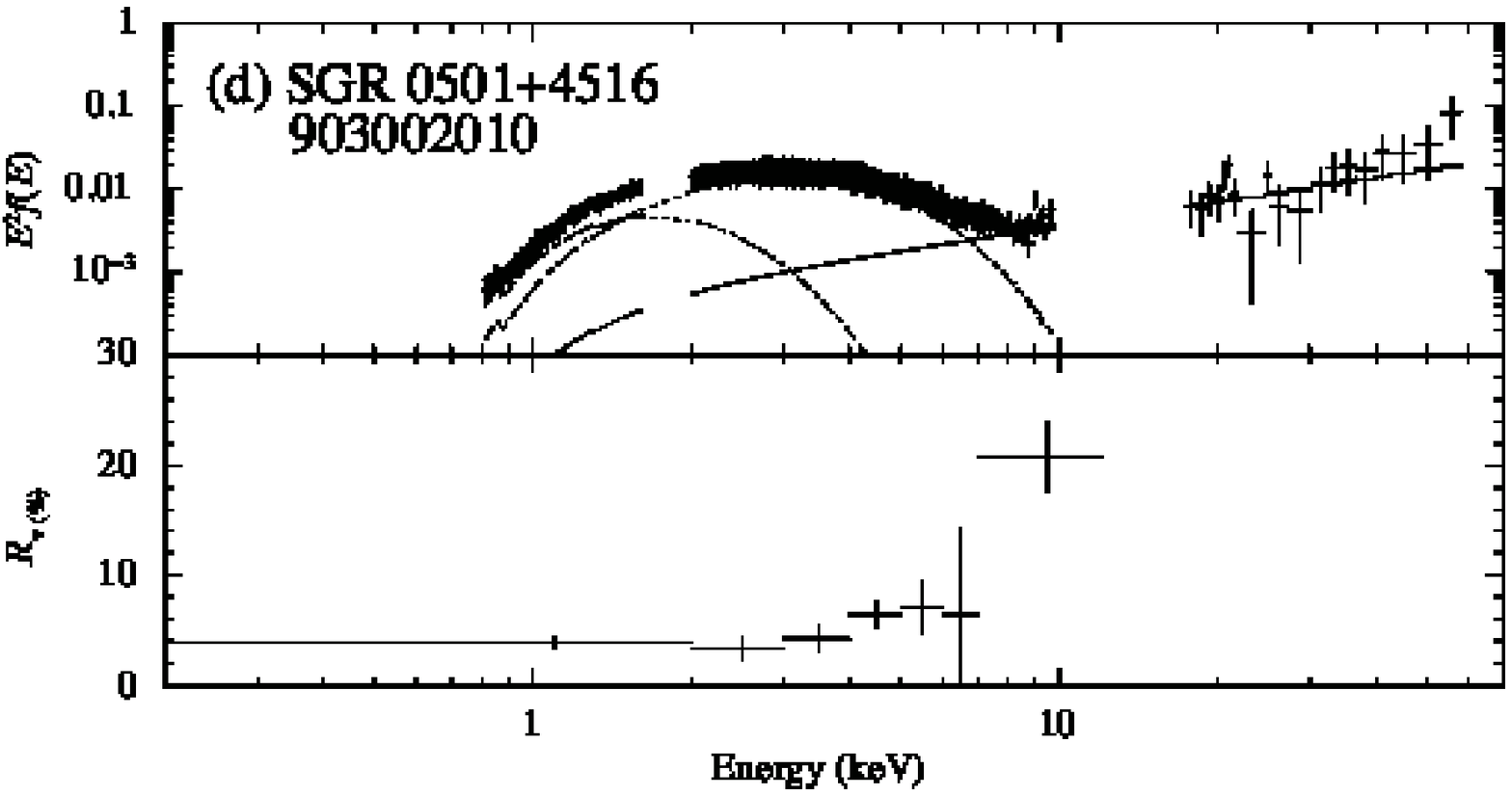}
   \end{minipage}
   \begin{minipage}{0.32\hsize}
   \includegraphics[height=2.7cm,width=5.46cm]{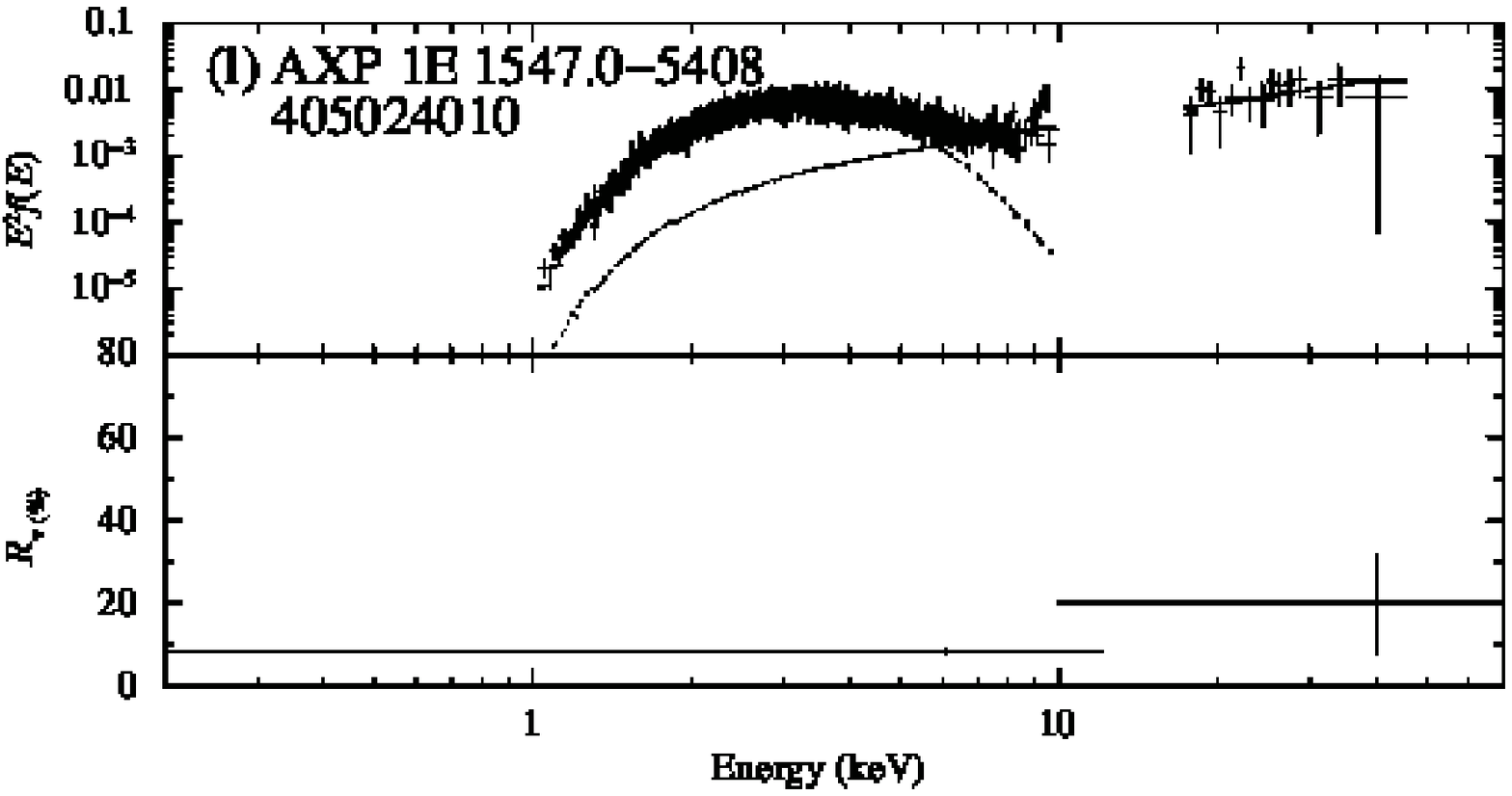}
   \end{minipage}
   \begin{minipage}{0.32\hsize}
  \includegraphics[height=2.7cm,width=5.46cm]{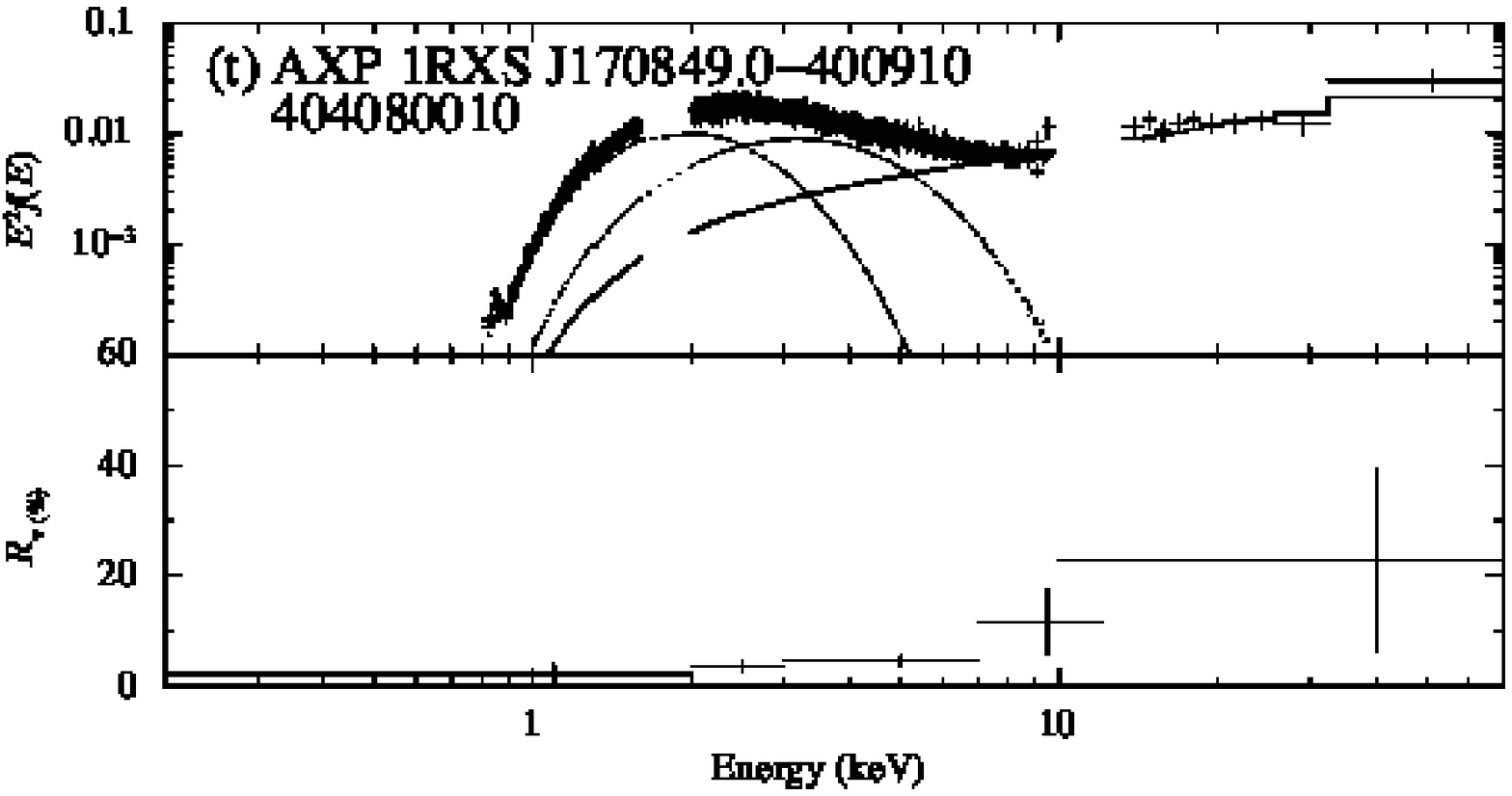}
   \end{minipage}
   \\
   \begin{minipage}{0.32\hsize}
    \includegraphics[height=2.7cm,width=5.46cm]{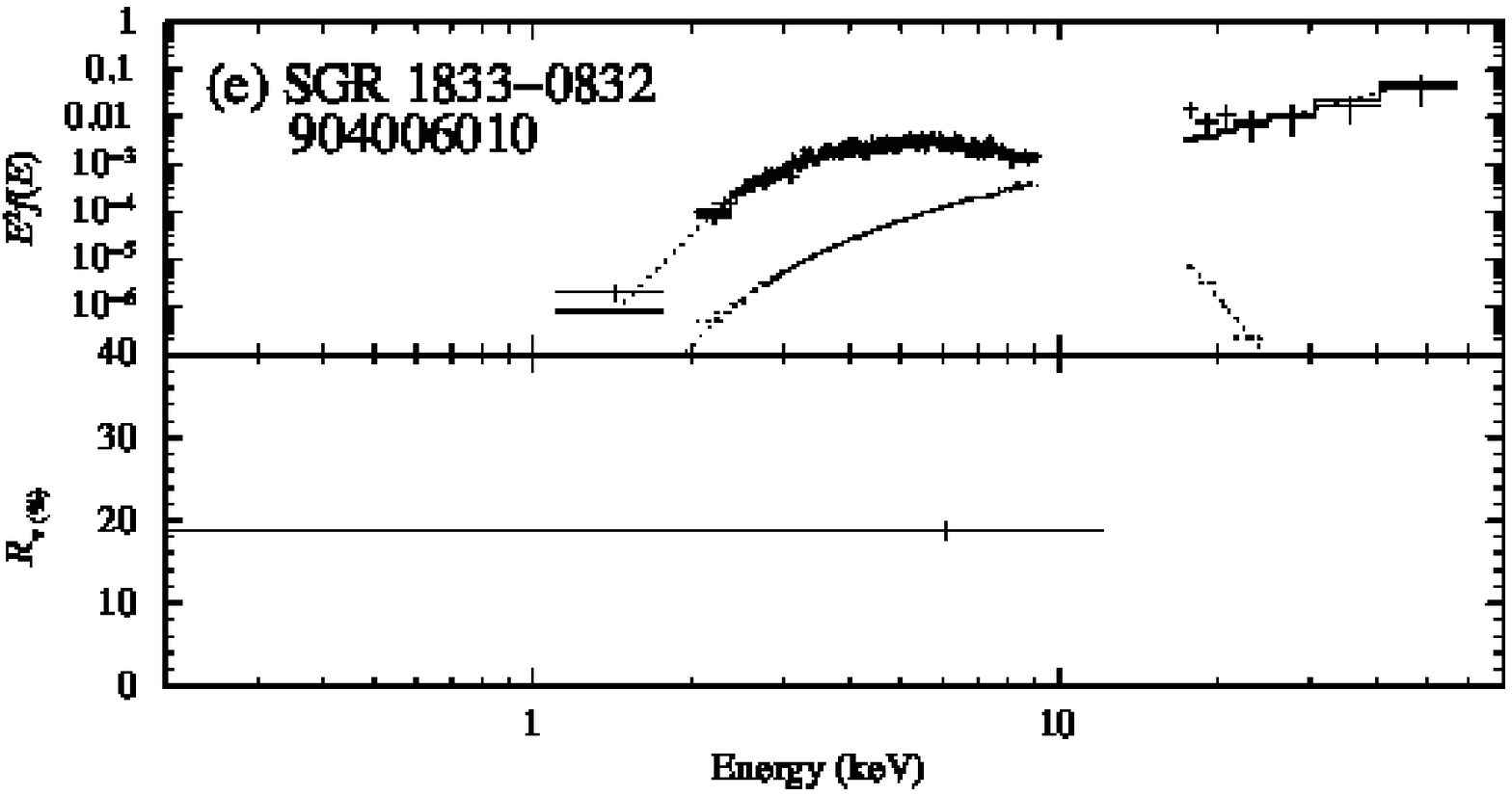}
   \end{minipage}
   \begin{minipage}{0.32\hsize}
    \includegraphics[height=2.7cm,width=5.46cm]{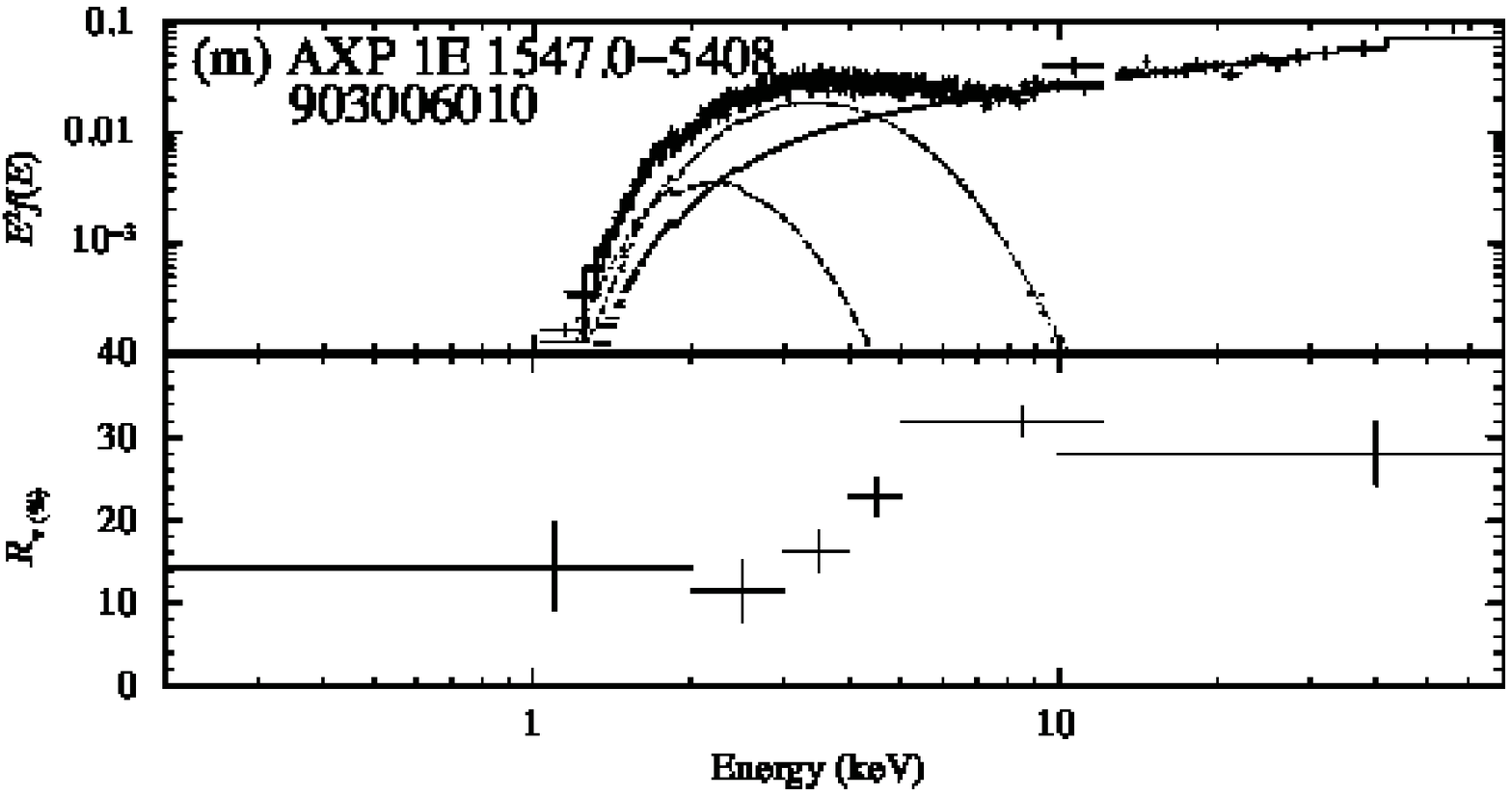}
   \end{minipage}
   \begin{minipage}{0.32\hsize}
    \includegraphics[height=2.7cm,width=5.46cm]{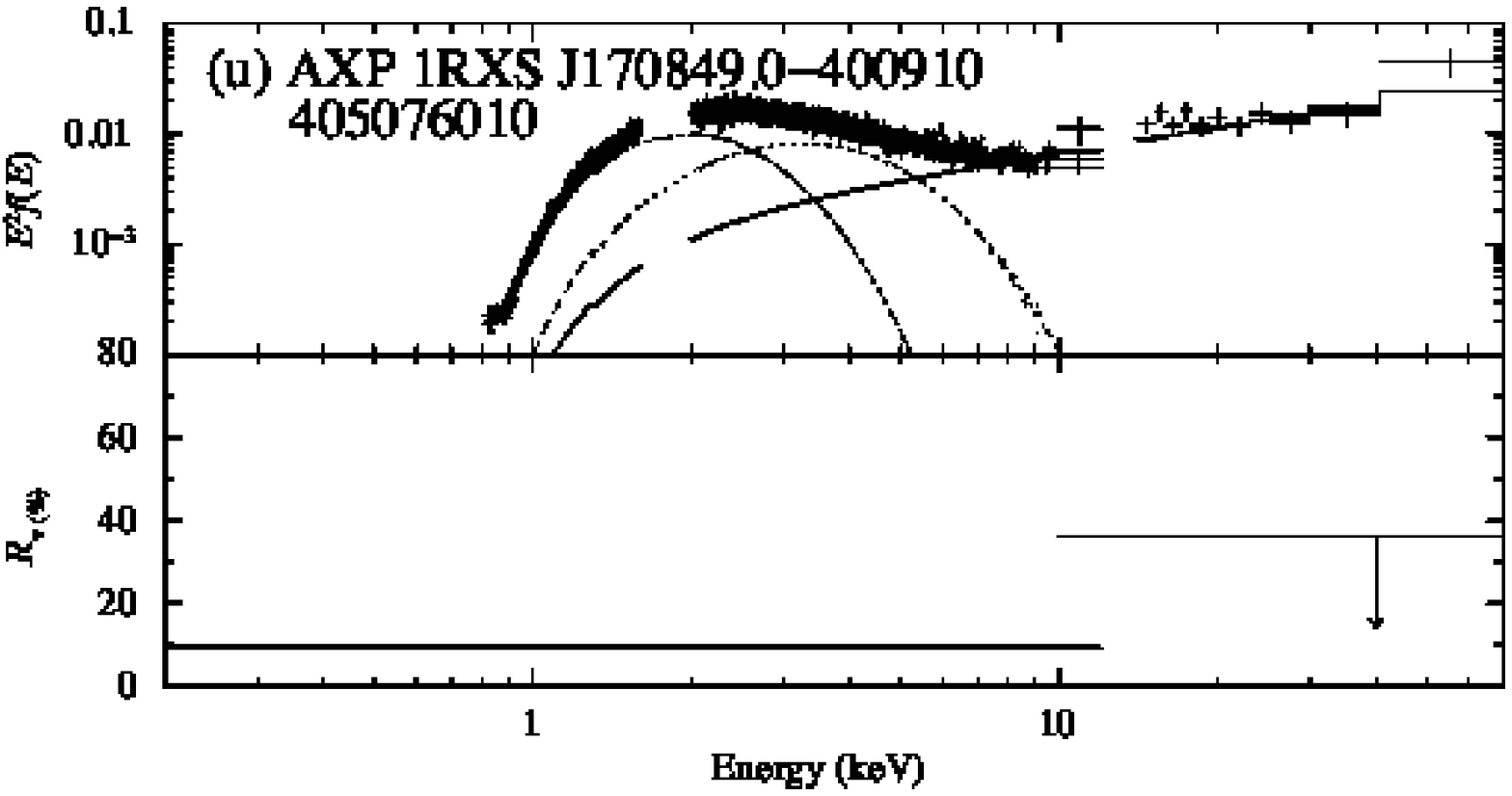}
   \end{minipage}
   \\
   \begin{minipage}{0.32\hsize}
     \includegraphics[height=2.7cm,width=5.46cm]{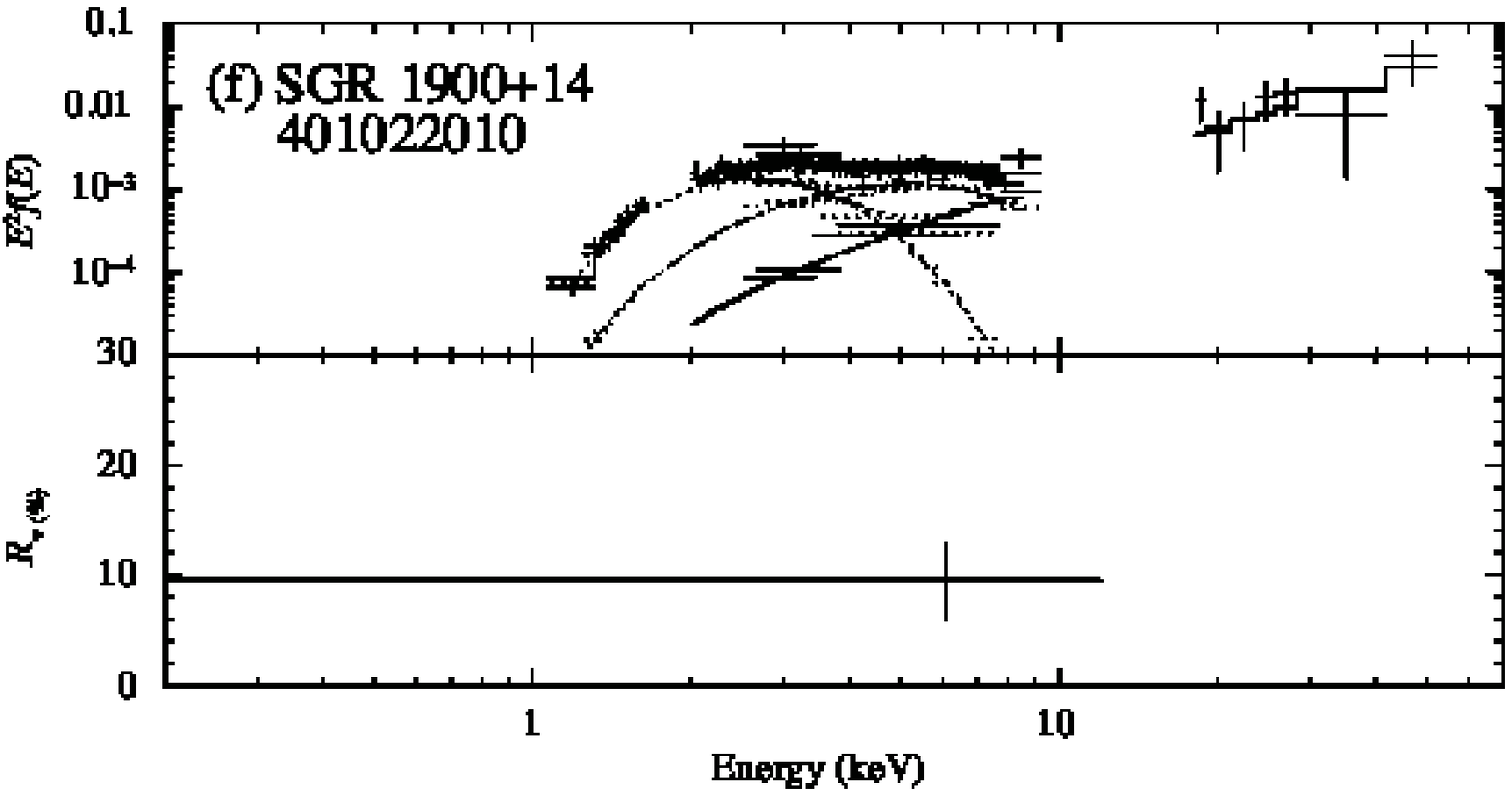}
   \end{minipage}
   \begin{minipage}{0.32\hsize}
    \includegraphics[height=2.7cm,width=5.46cm]{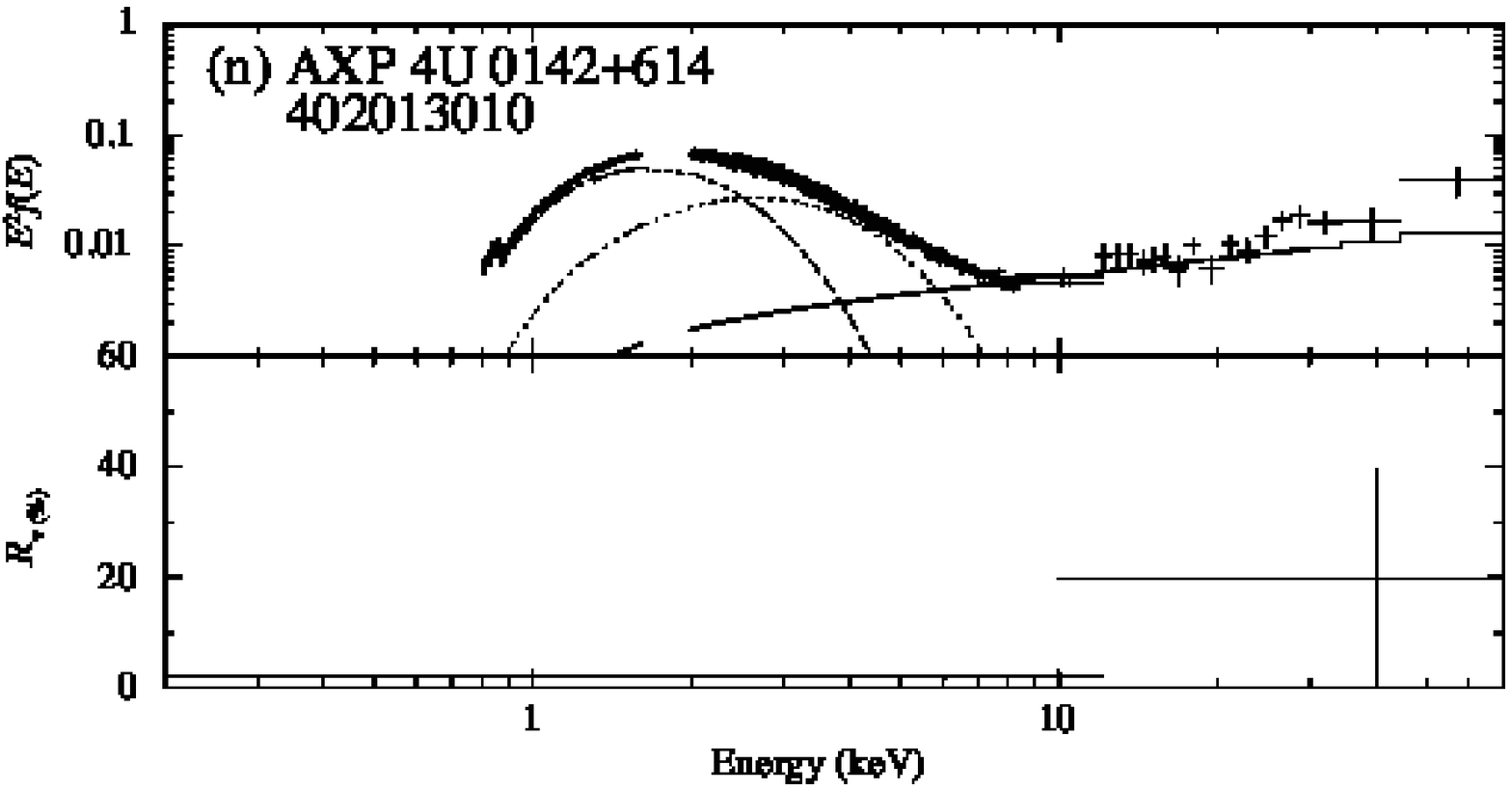}
   \end{minipage}
   \begin{minipage}{0.32\hsize}
        \includegraphics[height=2.7cm,width=5.46cm]{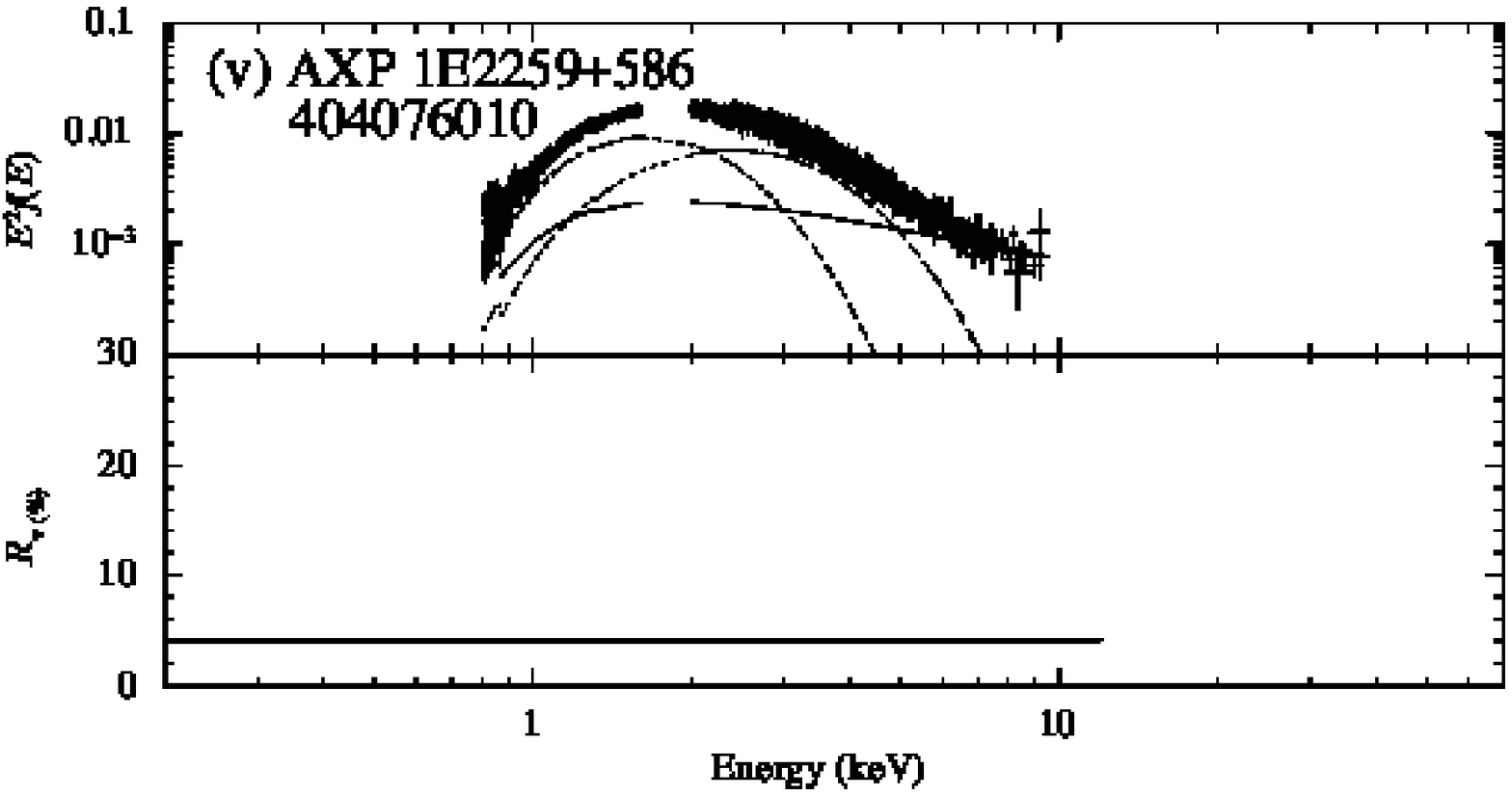}
   \end{minipage}
   \\
   \begin{minipage}{0.32\hsize}
   \includegraphics[height=2.7cm,width=5.46cm]{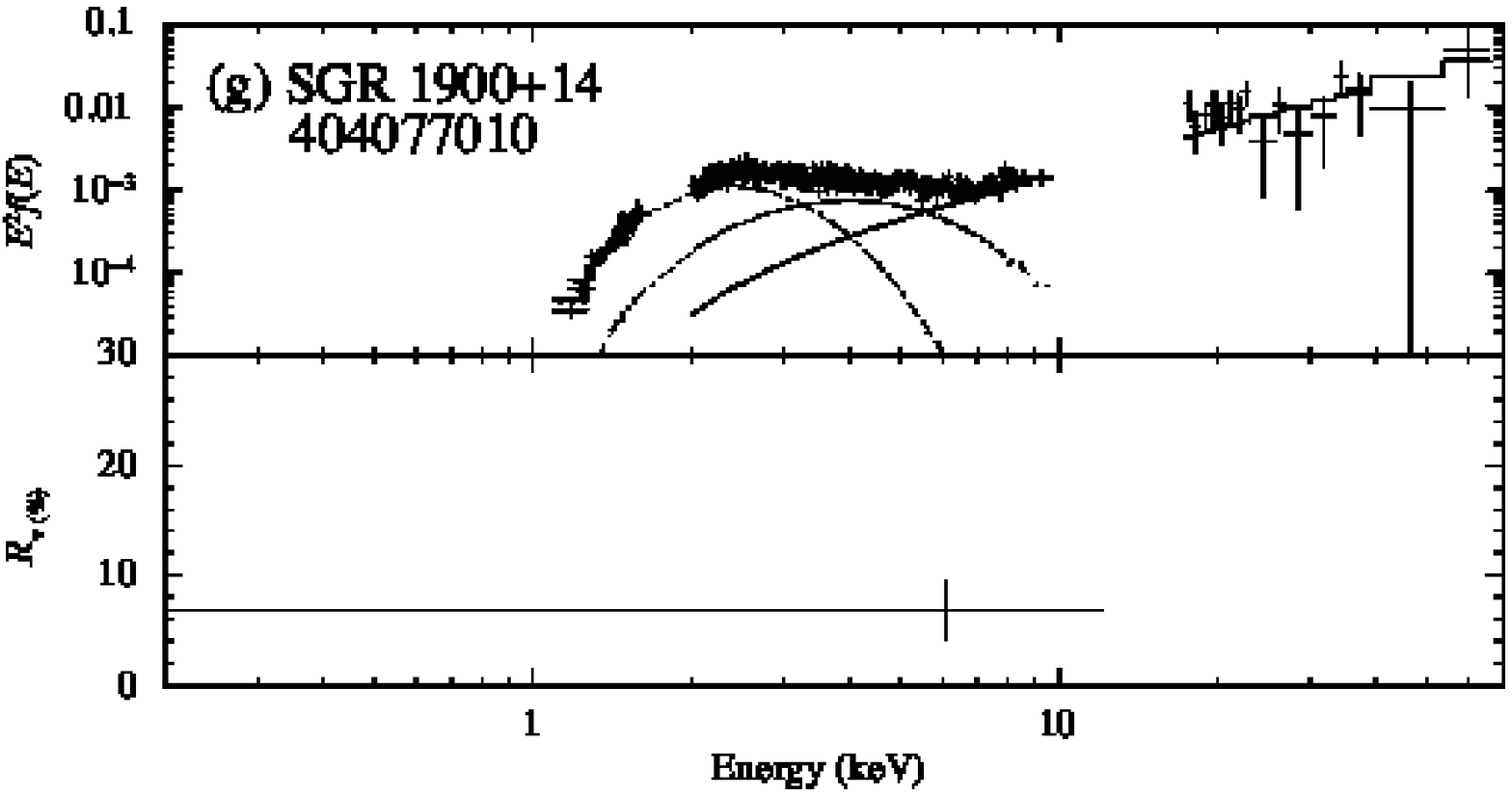}
   \end{minipage}
   \begin{minipage}{0.32\hsize}
      \includegraphics[height=2.7cm,width=5.46cm]{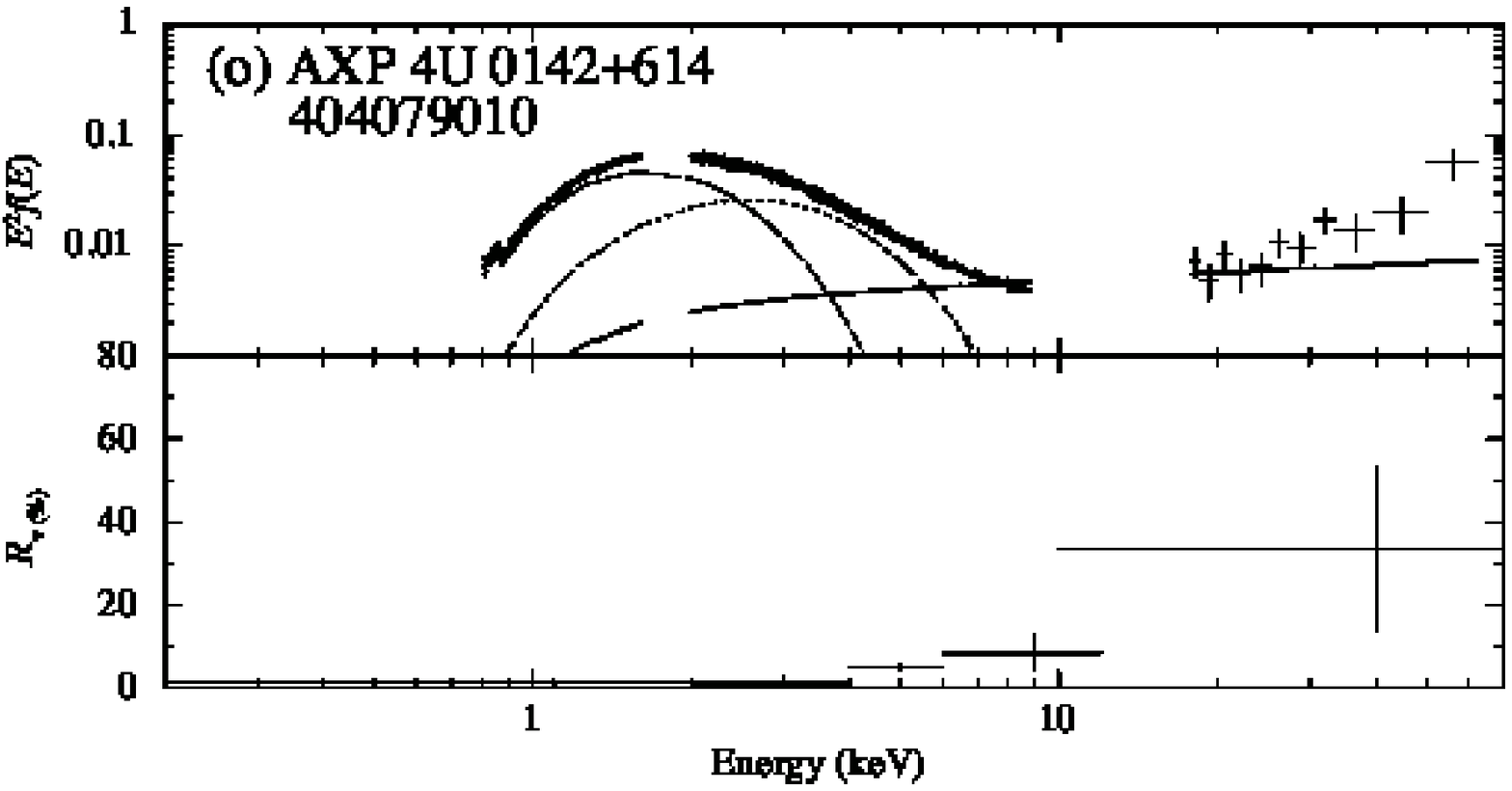}
   \end{minipage}
   \begin{minipage}{0.32\hsize}
 	\hspace{5cm}
   \end{minipage}
    \\
   \begin{minipage}{0.32\hsize}
    \includegraphics[height=2.7cm,width=5.46cm]{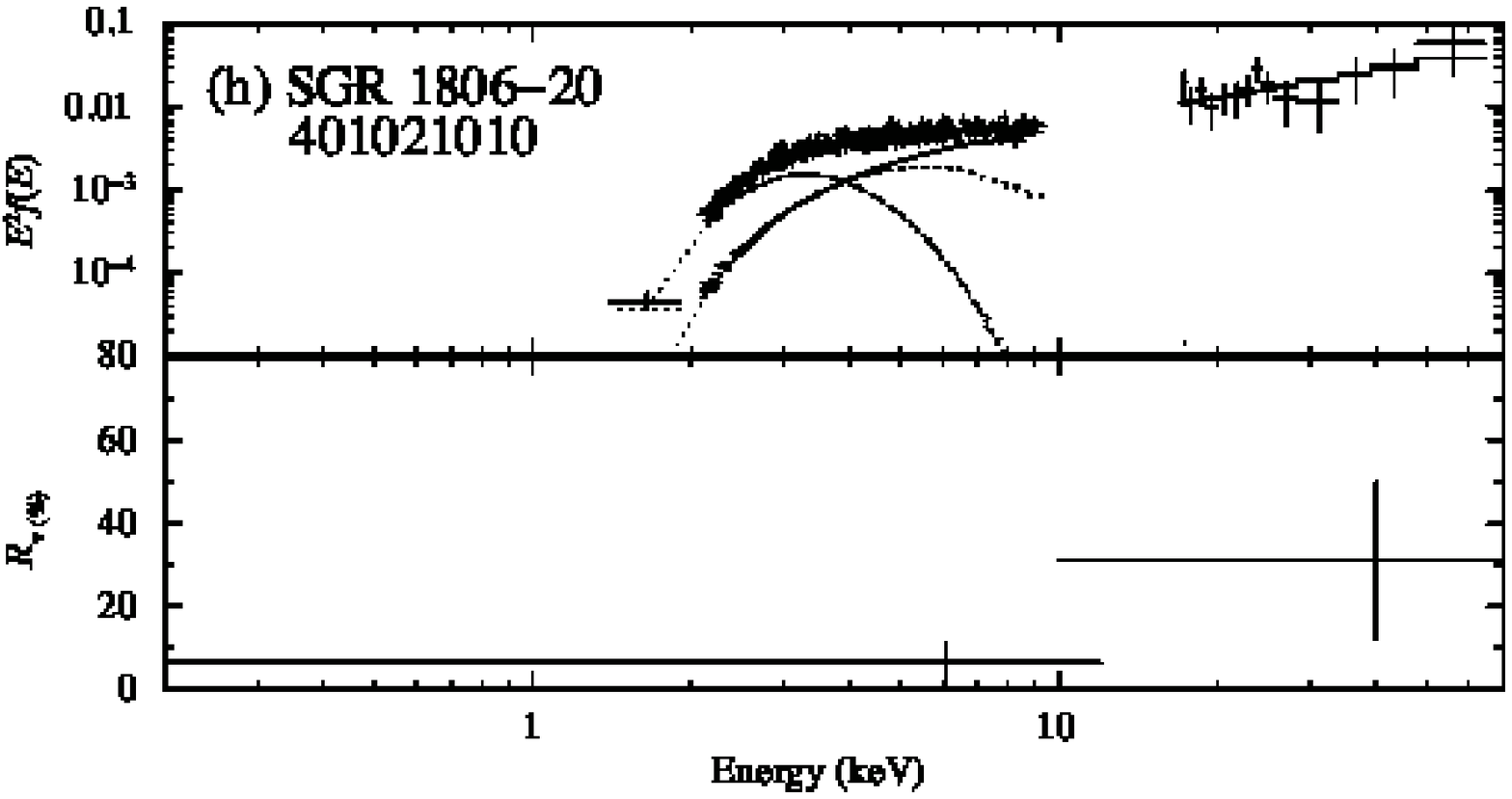}
   \end{minipage}
   \begin{minipage}{0.32\hsize}
 	\includegraphics[height=2.7cm,width=5.46cm]{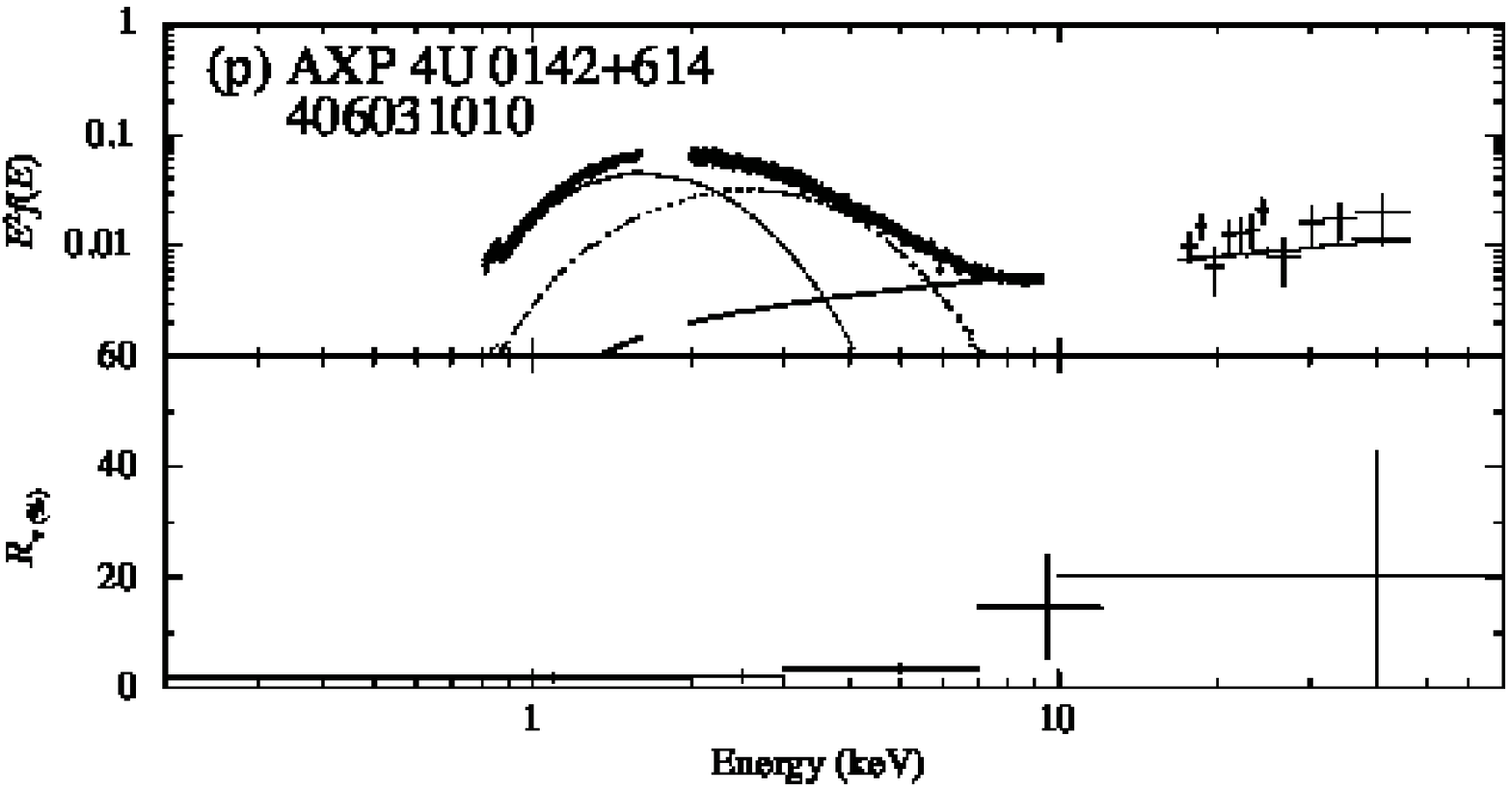}
   \end{minipage}
   \begin{minipage}{0.32\hsize}
 	\hspace{5cm}
   \end{minipage}
   \end{center}
  \caption{
  $E^2{f}(E)$ spectra, and spectra of the RMS intensity variations.
  The spectra in the panels (c), (q), (r), (s) and (v) are fitted with 2BB.
  The spectrum in the panel (e) is fitted with BB$+$PL.
  The spectra in the other panels are fitted with 2BB$+$PL.
  }
  \label{fig:rms_spc_summary}
\end{figure*}

We  discovered that energy dependencies of the RMS intensity variations
for 4 magnetars (6 observations;  panels (a), (b), (d), (h), (m) and (o) in figure \ref{fig:rms_spc_summary}),
clearly increase toward higher energy bands.
Among these observations, the RMS intensity variations remarkably increase above $\sim$8\,keV
and $\sim$4\,keV for SGR\,0501$+$4516 (OBSID=404078010;  panel (a) in figure \ref{fig:rms_spc_summary})
and AXP\,1E\,1547.0$-$5408 (OBSID=903006010; panel (m) in figure \ref{fig:rms_spc_summary}), respectively.
These energies correspond to the crossing points of the soft thermal components and the hard X-ray components,
suggesting that the most variation is associated with the hard components.

The energy dependent RMS intensity variations may be explained by the micro-burst  model presented
in section \ref{sec:math_rms}.
It is reported that indices of the  cumulative number-intensity distribution  of short bursts increase
toward  higher energy bands  in SGR 1806--20 \citep{nakagawa2007}.
Increase of the index causes a high dispersion of fluences, 
which leads to a large RMS intensity variation as shown in figure \ref{fig:relation_alpha_rms} (right).
If the same energy dependence of the index is applicable to the micro-bursts,
smaller RMS intensity variation in  lower energy bands  is  caused by  smaller indices, and vice versa.

\subsection{Comparison with  Theoretical Models}
The present  results  will give constraints on persistent X-ray emission mechanisms of the magnetars. 
Significant RMS intensity variations in both the 0.2--12\,keV energy band and the 10--70\,keV energy band
imply that neither  the soft thermal component nor  the hard X-ray component is 
from the stable neutron star surface in thermal equilibrium.
In this context, thermal bremsstrahlung model at the neutron star surface \citep{thompson2005, beloborodov2007}
 is unlikely.  

Energy dependence of the RMS intensity variations suggests that 
the emission regions of the soft thermal component and  the hard X-ray component
are located separately.
In the magnetospheric synchrotron model  \citep{heyl2005b},
the soft thermal component is emitted from a fireball near the neutron star
and the hard X-ray component is emitted via synchrotron process 
far from the neutron star.
Thus, this model seems consistent with the present  result of the energy dependent RMS intensity variations.
In fact, power-law indices of the hard X-ray components expected from the synchrotron  model
(0.5; \cite{heyl2005b}) is comparable to the observed indices (0.3--1.7; \cite{enoto2010c}).
Furthermore, the synchrotron model is applicable not only to the persistent emission but also to the burst emissions \citep{heyl2005a}.
This agrees with our idea that the persistent X-ray emission is composed of  numerous micro-bursts, and that the
persistent emission and the bust emission have the same origin.

\subsection{A Unified View of the Magnetar X-ray Emission}
\begin{figure}
  \begin{center}
     \includegraphics[angle=0,width=16.38cm]{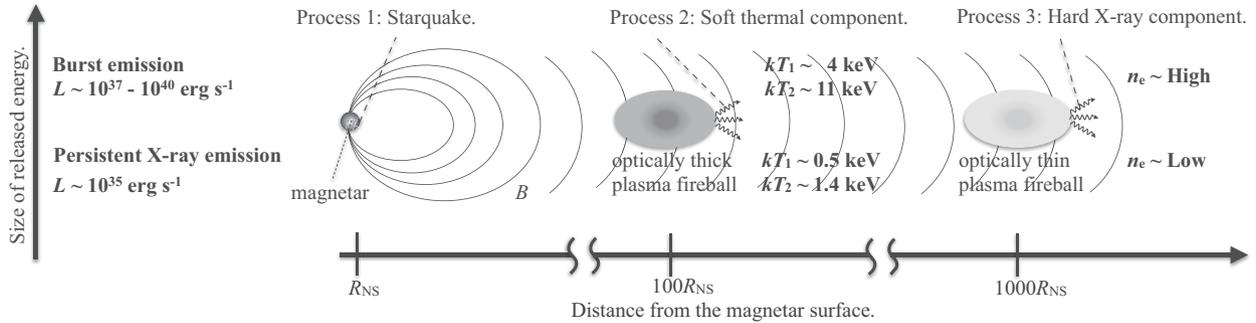}
  \end{center}
  \caption{
  A schematic illustration of our "Mircro-Burst Model"
  of both the persistent X-ray emission and the burst emission from magnetars.
  $R_{\rm NS}$ indicates a typical neutron star radius of $\sim$10\,km, $L$ indicates a typical luminosity,
  $kT_{\rm 1}$ and $kT_{\rm 2}$ indicate typical blackbody temperatures and $n_{\rm e}$ indicates
  number of electrons in emission regions.
  Magnetic disturbance may be strong at a distance of $\sim$1000$R_{\rm NS}$ and may cause
  the large RMS intensity variations.
  }
  \label{fig:magnetar_view}
\end{figure}

Figure \ref{fig:magnetar_view} shows a schematic illustration of our unified view
("Micro-Burst Model") of both the persistent X-ray emission and the burst emission from
magnetars,  based on the present observation and 
theoretical models \citep{duncan1992, thompson1995, heyl2005a, heyl2005b}.
In this  model, the burst emission is  caused by a single energetic
fireball, while the persistent X-ray emission consists of  numerous micro-bursts
 caused by numerous small fireballs. 
In the unified view, both the persistent X-ray emission and the burst emission are explained under
the same configuration, where only
their luminosities are different.
The persistent X-ray emission and the burst emission have
typical luminosities of $\sim$10$^{35}$\,erg\,s$^{-1}$ and $\sim$10$^{37}$ -- $\sim$10$^{40}$\,erg\,s$^{-1}$, respectively (e.g., \cite{nakagawa2011a}).

Initially, a starquake occurs on the  magnetar surface (\cite{duncan1992}; process 1 in figure \ref{fig:magnetar_view}).
The starquake produces an electron-positron pair plasma fireball which has  momenta
to leave from the magnetar.
The fireball travels in the magnetosphere and emits blackbody
emissions (i.e., the soft-thermal components)
at around $\sim100R_{\rm NS}$
where  $R_{\rm NS}$ is a  typical neutron star radius of $\sim$10\,km (\cite{heyl2005b}; process 2 in figure \ref{fig:magnetar_view}).
The  emissions is 
observed as  two blackbody spectra,
because of two different polarization modes in strong magnetic fields
of $\sim10^{14}$\,G \citep{thompson1995}.
Their typical temperatures are  $\sim$0.5\,keV and $\sim$1.4\,keV for the persistent X-ray emission (e.g., \cite{nakagawa2009a}),
or  $\sim$4\,keV and $\sim$11\,keV for the burst emission (e.g., \cite{nakagawa2007}).

Eventually, the fireball turns to optically thin condition and emits synchrotron emissions (i.e., the hard X-ray components)
at around $\sim1000R_{\rm NS}$ (\cite{heyl2005b}; process 3 in figure \ref{fig:magnetar_view}).
The spatial scale of causality for the hard X-ray components is estimated to be $\sim30R_{\rm NS}$
in assuming that the micro-burst has 1\,ms duration.
Therefore the size of each fireball should be less than $\sim30R_{\rm NS}$.
Number of the electrons 
to produce the hard X-ray component for the persistent X-ray emission
may be lower by 2--3 orders of magnitude than that for the burst
emission  \citep{nakagawa2011a}.
Presumably, the  magnetic disturbance is increased outward from the magnetars,
so that the strong magnetic disturbance causes the larger  RMS
intensity variations in the hard X-ray components.

\section{Conclusion}
Using the {\it Suzaku} archive data of  11 magnetars (22 observations),
we found significant RMS intensity variations in the persistent X-ray emission
from all the magnetars studied.
In addition, we found that the RMS intensity variations increase
toward higher energy band for 4 magnetars (6 observations).
These RMS intensity variations are consistent with the
micro-burst model, where  
the persistent X-ray emission is composed of
numerous mircro-bursts of various sizes subject to a particular cumulative number-intensity distribution.
We propose a unified view of the magnetar X-ray emission based on 
the present  results.

Time scale of the flux variations expected in the micro-burst model 
is  shorter than a few milliseconds \citep{nakagawa2007}.
Future observations of magnetars  with a higher sensitivity and
better time resolution, which is expected to give more
accurate measurements of the RMS intensity variations
will allow us to strongly constrain  magnetar emission mechanisms
by directly comparing with theoretical models (e.g., \cite{duncan1992, thompson1995, heyl2005a, heyl2005b}).

%
{\setlength{\tabcolsep}{2.5pt}
\begin{longtable}{lllccccccc}
  \caption{Summary of magnetars used in this study and their RMS intensity variations.}\label{tab:suzaku_obs_list}
      \hline
      Object\footnotemark[$*$]                & OBSID\footnotemark[$\dagger$] & Date\footnotemark[$\ddagger$] & $T_{\mathrm{X}}${\footnotemark[$\S$]} & $T_{\mathrm{P}}${\footnotemark[$\S$]} & $N_{b}${\footnotemark[$\|$]} &
      $R_{\mathrm{X}}${\footnotemark[$\#$]} & $R_{\mathrm{P}}${\footnotemark[$**$]}  & $R'_{\mathrm{X}}${\footnotemark[$\dagger\dagger$]} & $R'_{\mathrm{P}}${\footnotemark[$\ddagger\ddagger$]} \\
                                    &               & (UTC)        & (ks)                           & (ks)                               &    & (\%)                            & (\%) & (\%)                            & (\%) \\
      \endfirsthead
      \hline
      Object\footnotemark[$*$]                & OBSID\footnotemark[$\dagger$] & Date\footnotemark[$\ddagger$] & $T_{\mathrm{X}}${\footnotemark[$\S$]} & $T_{\mathrm{P}}${\footnotemark[$\S$]} & $N_{b}${\footnotemark[$\|$]} &
      $R_{\mathrm{X}}${\footnotemark[$\#$]} & $R_{\mathrm{P}}${\footnotemark[$**$]}  & $R'_{\mathrm{X}}${\footnotemark[$\dagger\dagger$]} & $R'_{\mathrm{P}}${\footnotemark[$\ddagger\ddagger$]} \\
                                    &               & (UTC)        & (ks)                           & (ks)                               &    & (\%)                            & (\%) & (\%)                            & (\%) \\
      \endhead
      \hline
    \multicolumn{10}{l}{\footnotemark[$*$] Object name of the SGRs and AXPs.} \\
	\multicolumn{10}{l}{\footnotemark[$\dagger$] {\it Suzaku} observation ID.} \\
	\multicolumn{10}{l}{\footnotemark[$\ddagger$] Date of the observation start.} \\
	\multicolumn{10}{l}{\footnotemark[$\S$] Net exposures of the XIS ($T_{\mathrm{X}}$) and the HXD-PIN ($T_{\mathrm{P}}$).} \\
	\multicolumn{10}{l}{\footnotemark[$\|$] Number of bins where photon counts exceed a burst criteria. See details in text.} \\
	\multicolumn{10}{l}{\footnotemark[$\#$] The RMS intensity variations in 0.2--12\,keV (XIS).} \\
	\multicolumn{10}{l}{\footnotemark[$**$] The RMS intensity variations in 10--70\,keV (HXD-PIN).} \\
	\multicolumn{10}{l}{\footnotemark[$\dagger\dagger$] The RMS intensity variations without effect of the burst emission in 0.2--12\,keV (XIS).} \\
	\multicolumn{10}{l}{\footnotemark[$\ddagger\ddagger$] The RMS intensity variations without effect of the burst emission in 10--70\,keV (HXD-PIN).} \\
	\multicolumn{10}{l}{\footnotemark[$\S\S$] The HXD-PIN data did not show a statistically significant signal.} \\
	\multicolumn{10}{l}{\footnotemark[$\|\|$] The HXD-PIN data was not used due to a contamination source in a field of view.} \\
      \endfoot
      \hline
    \multicolumn{10}{l}{\footnotemark[$*$] Object name of the SGRs and AXPs.} \\
	\multicolumn{10}{l}{\footnotemark[$\dagger$] {\it Suzaku} observation ID.} \\
	\multicolumn{10}{l}{\footnotemark[$\ddagger$] Date of the observation start.} \\
	\multicolumn{10}{l}{\footnotemark[$\S$] Net exposures of the XIS ($T_{\mathrm{X}}$) and the HXD-PIN ($T_{\mathrm{P}}$).} \\
	\multicolumn{10}{l}{\footnotemark[$\|$] Number of bins where photon counts exceed a burst criteria. See details in text.} \\
	\multicolumn{10}{l}{\footnotemark[$\#$] The RMS intensity variations in 0.2--12\,keV (XIS).} \\
	\multicolumn{10}{l}{\footnotemark[$**$] The RMS intensity variations in 10--70\,keV (HXD-PIN).} \\
	\multicolumn{10}{l}{\footnotemark[$\dagger\dagger$] The RMS intensity variations without effect of the burst emission in 0.2--12\,keV (XIS).} \\
	\multicolumn{10}{l}{\footnotemark[$\ddagger\ddagger$] The RMS intensity variations without effect of the burst emission in 10--70\,keV (HXD-PIN).} \\
	\multicolumn{10}{l}{\footnotemark[$\S\S$] The HXD-PIN data did not show a statistically significant signal.} \\
	\multicolumn{10}{l}{\footnotemark[$\|\|$] The HXD-PIN data was not used due to a contamination source in a field of view.} \\
      \endlastfoot
      \hline
      SGR\,0501$+$4516              & 404078010 & 2009-08-17 & 43 & 25 & 0 &  9.2$\pm$1.1   &   $50\pm6$  & 9.2$\pm$1.1 & $50\pm6$ \\
                                    & 405075010 & 2010-09-20 & 60 & 49 & 0 &   7.2$\pm$2.5  &  $99\pm86$   & 7.2$\pm$2.5 & $99\pm86$ \\
                                    & 408013010\footnotemark[$\S\S$] & 2013-08-31 & 41 & 33 & 0 &  14$\pm$1   &   $-$  & 14$\pm$1 & $-$ \\
                                    & 903002010 & 2008-08-26 & 60 & 50 & 17 &   50.9$\pm$0.5  &  0   & 7.37$\pm$0.07 & 0 \\
      SGR\,1833$-$0832              & 904006010 & 2010-03-27 & 42 & 10 & 2 &   27.5$\pm$0.7  &  0   & 18.8$\pm$1.1 & 0 \\
      SGR\,1900$+$14                & 401022010 & 2006-04-01 & 22 & 13 & 0 &   6.8$\pm$2.7  &  0   & 6.8$\pm$2.7 & 0 \\
                                    & 404077010 & 2009-04-26 & 53 & 39 & 0 &   9.6$\pm$3.5  &  0  & 9.6$\pm$3.5 & 0 \\
      SGR\,1806$-$20                & 401021010 & 2007-03-30 & 19 & 16 & 2 &   51$\pm$1  &   26$\pm$24  & 6.4$\pm$4.8 & $31\pm19$ \\
                                    & 401092010 & 2006-09-09 & 49 & 52 & 9 &  135$\pm$1   &   59$\pm$5  & 6.5$\pm$0.8 & $17\pm13$ \\
                                    & 402094010 & 2007-10-14 & 52 & 46 & 6 &   82.7$\pm$0.9  &  75$\pm$7   & 4.2$\pm$0.5 & $<55$ \\
                                    & 406069010 & 2012-03-24 & 71 & 60 & 0 &   16.4$\pm$0.4  &   0  & 16.4$\pm$0.4 & 0 \\
      AXP\,1E\,1547.0$-$5408        & 405024010 & 2010-08-07 & 52 & 40 & 0 &   8.4$\pm$0.6  &   $20\pm12$  & 8.4$\pm$0.6 & $20\pm12$ \\
                                    & 903006010 & 2009-01-28 & 11 & 31 & 25 &   36.0$\pm$0.3  &  32$\pm$3   & 18.2$\pm$0.2 & $28\pm3$ \\
      AXP\,4U\,0142$+$614           & 402013010 & 2007-08-13 & 100 & 95 & 0 &  2.0$\pm$0.1   &   $<40$  & 2.0$\pm$0.1 & $<40$ \\
                                    & 404079010 & 2009-08-10 & 107 & 92 & 0 &   1.3$\pm$0.3  &   $33\pm20$  & 1.3$\pm$0.3 & $33\pm20$ \\
                                    & 406031010 & 2011-09-07 & 39 & 39 & 0 &   1.8$\pm$0.3  &   $<42$  & 1.8$\pm$0.3 & $<42$ \\
      AXP\,1E\,1048.1$-$5937        & 403005010\footnotemark[$\S\S$] & 2008-11-30 & 100 & 63 & 1 &  3.0$\pm$1.0   &   $-$  & 3.0$\pm$1.0 & $-$ \\
      AXP\,Swift\,J1822.3$-$1606    & 906002010\footnotemark[$\S\S$] & 2011-09-13 & 41 & 34 & 0 &  6.0$\pm$0.3   &  $-$   & 6.0$\pm$0.3 & $-$ \\
      AXP\,CXOU\,J164710.2$-$455216 & 901002010\footnotemark[$\|\|$] & 2006-09-23 & 39 & $-$ & 0 &   3.3$\pm$0.7  &  $-$    & 3.3$\pm$0.7 & $-$ \\
      AXP\,1RXS\,J170849.0$-$400910 & 404080010 & 2009-08-23 & 61 & 48 & 0 &    7.85$\pm$0.07 &   $23\pm16$  & 7.85$\pm$0.07 & $23\pm16$ \\
                                    & 405076010 & 2010-09-27 & 63 & 55 & 0 &   9.25$\pm$0.06  &   $<36$  & 9.25$\pm$0.06 & $<36$ \\
      AXP\,1E\,2259$+$586           & 404076010\footnotemark[$\S\S$] & 2009-05-25 & 123 & 96 & 0 &   4.1$\pm$0.1  &  $-$    & 4.1$\pm$0.1 & $-$ \\
\end{longtable}
}


\begin{ack}
This work was supported by JSPS KAKENHI Grant Numbers 24540309, 15K05117 and 16K05309.
\end{ack}

\clearpage
\appendix 
\section*{RMS Intensity Variations Caused by Background Fluctuations}
Suzaku data of hard and bright non-variable sources (table \ref{tab:suzaku_obs_list_bkg}) are analyzed, 
in order to estimate the RMS intensity variations caused by background fluctuations and their time dependencies.
The RMS intensity variations are found to be $R_{\mathrm{X}}=$0.8--2.5\% in the 0.2--12\,keV energy band
and $R_{\mathrm{P}}=$5--12\% in the 10--70\,keV energy band, depending on observation periods.
We found that the time dependency is marginal
and confirmed that the background fluctuations are not significant for most cases.

{\setlength{\tabcolsep}{1.5pt}
\begin{longtable}{lllcccc}
  \caption{Summary of objects used for the background fluctuations and their RMS intensity variations.}\label{tab:suzaku_obs_list_bkg}
      \hline
      Object\footnotemark[$*$]                & OBSID\footnotemark[$\dagger$] & Date\footnotemark[$\ddagger$] & $T_{\mathrm{X}}${\footnotemark[$\S$]} & $T_{\mathrm{P}}${\footnotemark[$\S$]} &
      $R_{\mathrm{X}}${\footnotemark[$\|$]} & $R_{\mathrm{P}}${\footnotemark[$\#$]}  \\
                                    &               & (UTC)        & (ks)                           & (ks)                                  & (\%)                            & (\%) \\
      \endfirsthead
      \hline
      Object\footnotemark[$*$]                & OBSID\footnotemark[$\dagger$] & Date\footnotemark[$\ddagger$] & $T_{\mathrm{X}}${\footnotemark[$\S$]} & $T_{\mathrm{P}}${\footnotemark[$\S$]} &
      $R_{\mathrm{X}}${\footnotemark[$\|$]} & $R_{\mathrm{P}}${\footnotemark[$\#$]}  \\
                                    &               & (UTC)        & (ks)                           & (ks)                                  & (\%)                            & (\%) \\
      \endhead
      \hline
        \multicolumn{7}{l}{\footnotemark[$*$] Object name.} \\
	\multicolumn{7}{l}{\footnotemark[$\dagger$] {\it Suzaku} observation ID.} \\
	\multicolumn{7}{l}{\footnotemark[$\ddagger$] Date of the observation start.} \\
	\multicolumn{7}{l}{\footnotemark[$\S$] Net exposures of the XIS ($T_{\mathrm{X}}$) and the HXD-PIN ($T_{\mathrm{P}}$).} \\
	\multicolumn{7}{l}{\footnotemark[$\|$] The RMS intensity variations in 0.2--12\,keV (XIS).} \\
	\multicolumn{7}{l}{\footnotemark[$\#$] The RMS intensity variations in 10--70\,keV (HXD-PIN).} \\
	\multicolumn{7}{l}{\footnotemark[$**$] The XIS data was not used due to a telemetry saturation \citep{maeda2009}.} \\
	\multicolumn{7}{l}{\footnotemark[$\dagger\dagger$] 
	\parbox[t]{50em}{\strut
	The data of  OBSID=508011020 for Cas\,A and OBSID=101012010 for Perseus\,Cluster were not used due to visually obvious flux variations in light curves.
	\strut}
	} \\
	\multicolumn{7}{l}{\footnotemark[$\ddagger\ddagger$] The XIS data was not available.} \\
	\multicolumn{7}{l}{\footnotemark[$\S\S$] The HXD-PIN data was not available.} \\
      \endfoot
      \hline
        \multicolumn{7}{l}{\footnotemark[$*$] Object name.} \\
	\multicolumn{7}{l}{\footnotemark[$\dagger$] {\it Suzaku} observation ID.} \\
	\multicolumn{7}{l}{\footnotemark[$\ddagger$] Date of the observation start.} \\
	\multicolumn{7}{l}{\footnotemark[$\S$] Net exposures of the XIS ($T_{\mathrm{X}}$) and the HXD-PIN ($T_{\mathrm{P}}$).} \\
	\multicolumn{7}{l}{\footnotemark[$\|$] The RMS intensity variations in 0.2--12\,keV (XIS).} \\
	\multicolumn{7}{l}{\footnotemark[$\#$] The RMS intensity variations in 10--70\,keV (HXD-PIN).} \\
	\multicolumn{7}{l}{\footnotemark[$**$] The XIS data was not used due to a telemetry saturation \citep{maeda2009}.} \\
	\multicolumn{7}{l}{\footnotemark[$\dagger\dagger$] 
	\parbox[t]{50em}{\strut
	The data of  OBSID=508011020 for Cas\,A and OBSID=101012010 for Perseus\,Cluster were not used due to visually obvious flux variations in light curves.
	\strut}
	} \\
	\multicolumn{7}{l}{\footnotemark[$\ddagger\ddagger$] The XIS data was not available.} \\
	\multicolumn{7}{l}{\footnotemark[$\S\S$] The HXD-PIN data was not available.} \\
      \endlastfoot
      \hline
       Cas\,A\footnotemark[$\dagger\dagger$]                & 100016010\footnotemark[$**$] & 2005-09-01 & $-$ & 24 & $-$ & 12$\pm$5 \\ 
                                  & 100043010\footnotemark[$\ddagger\ddagger$] & 2006-02-02 & $-$ & 10 & $-$ & $<40$ \\
                                  & 100043020 & 2006-02-17 & 7   &   16  & 0.8$\pm$0.5 & $<26$ \\
                                  & 507038010 & 2012-12-20 & 102 & 118 & 1.56$\pm$0.02 & $<28$ \\
       Coma\,Cluster    & 801097010 & 2006-05-31 & 179 & 157 &0.8$\pm$0.3 & 5$\pm$4 \\
       Perseus\,Cluster\footnotemark[$\dagger\dagger$]                                  & 101012020 & 2007-02-05 &  44  & 42 & $<1.6$ & 6$\pm$5 \\
                                  & 102011010 & 2007-08-15 &  42  & 36 & 0.9$\pm$0.7 & 7$\pm$5 \\
                                  & 102012010 & 2008-02-27 &  42  & 62 & 0.9$\pm$0.7 & 7$\pm$3 \\
                                  & 103004010 & 2008-08-13 &  41  & 32  & 1.5$\pm$0.3 & 0 \\
                                  & 103004020 & 2009-02-11 &  50  & 45 & 1.2$\pm$0.4 & $<16$ \\
                                  & 103005010 & 2008-08-14 & 21   & 17 & $<12$ & $<386$ \\  
                                  & 103005020 & 2009-02-12 &  29 &  27 & $<6$ & 0 \\ 
                                  & 104018010 & 2009-08-26 &  41  & 37 & 1.1$\pm$0.5 & $<46$ \\
                                  & 104019010 & 2010-02-01 &  39  & 36 & 1.7$\pm$0.3 & $<173$ \\
                                  & 104020010 & 2009-08-27 &  55  & 52 & 1.2$\pm$0.3 & $<20$ \\ 
                                  & 104021010 & 2010-02-02 &  22  & 22 & $<4$ & $<31$ \\ 
                                  & 105009010 & 2010-08-09 &  34  & 38 & 0.9$\pm$0.8 & 0 \\
                                  & 105009020 & 2011-02-03 &  40  & 30 & 1.3$\pm$0.4 & $<16$ \\
                                  & 105010010 & 2010-08-10 &  27  & 34 & $<2$ & $<20$ \\ 
                                  & 105010020 & 2011-02-02 &  21  & 17 & $<3$ & $<21$ \\ 
                                  & 105027010 & 2011-02-22 &  46  & 42 & 1.7$\pm$0.2 & 11$\pm$3 \\
                                  & 105028010 & 2011-02-21 &  21  & 18 & 1.3$\pm$1.2 & 0 \\ 
                                  & 106005010 & 2011-07-27 &  41  & 37 & $<6.4$ & $<412$ \\
                                  & 106005020 & 2012-02-07 &  47  & 45 & 1.2$\pm$0.4 & 0 \\
                                  & 106006010 & 2011-07-26 &  40  & 35 & 0.9$\pm$0.8 & 0 \\
                                  & 106007010 & 2011-08-23 &  21   & 19 & 1.3$\pm$0.8 & 0 \\ 
                                  & 106007020 & 2012-02-08 &  21  & 21 & $<3$ & 10$\pm$7 \\ 
                                  & 106008010 & 2011-08-22 &  23  & 22 & $<2$ & $<19$ \\ 
                                  & 107005010 & 2012-08-20 &  41  & 39 & 0.9$\pm$0.6 & 12$\pm$3 \\
                                  & 107005020 & 2013-02-11 &  41  & 36 & 1.1$\pm$0.6 & 0 \\
                                  & 107006010 & 2012-08-20 &  24  & 20 & $<2$ & 9$\pm$7 \\ 
                                  & 107006020 & 2013-02-12 &  22  & 18 & 2.0$\pm$0.4 & 10$\pm$6 \\ 
                                  & 108005010 & 2013-08-15 &  41  & 41 & 1.2$\pm$0.4 & $<38$ \\
                                  & 108005020{\footnotemark[$\S\S$]} & 2014-02-05 &  38  & $-$ & 1.2$\pm$0.5 & $-$ \\
                                  & 108006010 & 2013-08-16 &  22  & 19 & 1.2$\pm$0.8 & $<20$ \\
                                  & 108006020{\footnotemark[$\S\S$]} & 2014-02-06 &  19  & $-$ & $<2$ & $-$ \\ 
                                  & 109005010{\footnotemark[$\S\S$]} & 2014-08-27 &  20  & $-$ & $<3.5$ & $-$ \\
                                  & 109005020{\footnotemark[$\S\S$]} & 2015-03-03 &  37  & $-$  & 1.4$\pm$0.3 & $-$  \\
                                  & 109006010{\footnotemark[$\S\S$]} & 2015-03-04 &  24   & $-$ & 2.5$\pm$0.4  & $-$ \\ 
                                  & 109007010{\footnotemark[$\S\S$]} & 2015-03-05 &  31  & $-$ & 1.2$\pm$0.8 & $-$ \\ 
\end{longtable}
}

\clearpage


\end{document}